\documentclass[12pt]{article}
\usepackage{graphicx}
\usepackage{amsmath, amssymb}

\newcommand{\mbf}[1]{\mathbf{#1}}

\renewcommand{\bar}[1]{\overline{#1}}

\def\Dslash{\raise.15ex\hbox{/}\kern-.7em D}
\def\Pslash{\raise.15ex\hbox{/}\kern-.7em P}

\begin {document}
\begin{flushright}
{\small
SLAC-PUB-14208\\
Jlab-PHY-10-1135\\
July 2010}
\end{flushright}

\vspace{20pt}

\centerline{\LARGE \bf The AdS/QCD Correspondence }
\centerline{\LARGE \bf and Exclusive Processes}

\vspace{20 pt}

\centerline{\bf {
Stanley J. Brodsky\footnote{Electronic address:
sjbth@slac.stanford.edu}$^{a,b}$,
Guy F. de T\'eramond\footnote{Electronic address:
gdt@asterix.crnet.cr}$^{c}$, and
Alexandre Deur\footnote{Electronic address:
deurpam@jlab.org}$^{d}$
}}

\vspace{5 pt}

{\centerline {$^{a}${SLAC National Accelerator Laboratory,
Stanford University, Stanford, CA 94309, USA}}

\vspace{2pt}

{\centerline {$^{b}${CP$^3$-Origins,
Southern Denmark University,
Odense, Denmark}}

\vspace{2pt}

{\centerline {$^{c}${Universidad de Costa Rica, San Jos\'e, Costa Rica}}

\vspace{2pt}

{\centerline {$^{d}${Thomas Jefferson National Accelerator Facility,
Newport News, VA 23606, USA}}

 \vspace{15pt}

\begin{abstract}
The AdS/CFT correspondence between theories in AdS space and
conformal field theories in physical space-time provides an
analytic, semi-classical, color-confining model for strongly-coupled QCD.
The soft-wall AdS/QCD model modified by a positive-sign dilaton metric  leads to a remarkable one-parameter description of nonperturbative hadron dynamics at zero quark mass, including a zero-mass pion and a Regge spectrum of linear trajectories with the same slope in orbital angular momentum $L$ and radial
quantum number $n$ for both mesons and baryons. One also predicts the form of the non-perturbative effective coupling $\alpha_s^{AdS}(Q)$ and its $\beta$-function
which agrees with the effective
coupling $\alpha_{g_1}$ extracted from  the Bjorken sum rule.
Light-front holography, which connects the fifth-dimensional coordinate of AdS space $z$ to an invariant impact separation variable $\zeta$,
allows one to compute the analytic form of the frame-independent
light-front wavefunctions, the fundamental entities which encode
hadron properties as well as  decay constants, form
factors, deeply virtual Compton scattering, exclusive heavy hadron
decays  and other exclusive scattering amplitudes.
One thus obtains a relativistic description of hadrons in QCD at the amplitude level
with dimensional counting for hard exclusive reactions at high momentum transfer.
As specific examples we discuss the
behavior of the pion and nucleon  form factors in the space-like and time-like regions.
We also review the phenomenology
of exclusive processes including some anomalous empirical results.

\end{abstract}



\vfill

\newpage

\parindent=1.5pc
\baselineskip=16pt

\setcounter{footnote}{0}

\section{Introduction}

Exclusive processes play a key role in quantum chromodynamics, testing the primary quark and gluon interactions of QCD and the structure of hadrons at the amplitude level.  Two basic pictures have emerged based on perturbative QCD  (pQCD) and nonperturbative AdS/QCD. In pQCD
hard scattering amplitudes at a high scale $Q^2 >> \Lambda^2_{\rm QCD}$  factorize as a convolution of gauge-invariant hadron distribution amplitudes $\phi_H(x_i,Q)$ with the underlying hard scattering quark-gluon subprocess amplitude $T_H$.
The leading power fall-off of the hard scattering amplitude follows from the conformal scaling of
the underlying hard-scattering amplitude: $T_H \sim 1/Q^{n-4}$, where $n$ is
the total number of fields (quarks, leptons, or gauge fields)
participating in the hard
scattering~\cite{Brodsky:1973kr, Matveev:1973ra}. Thus the reaction
is dominated by subprocesses and Fock states involving the minimum
number of interacting fields.  In the case of $2 \to 2$ scattering
processes, this implies the dimensional counting rules $ {d\sigma/ dt}(A B \to C D) ={F_{A B \to C
D}(t/s)/ s^{n-2}},$ where $n = N_A + N_B + N_C +N_D$ and
$N_H$
is the minimum number of constituents of $H$.   The result is modified by the 
ERBL evolution~\cite{Lepage:1979zb,Efremov:1979qk} of the distribution amplitudes and the running of the QCD coupling.

It is striking that the dimensional counting rules are also a key feature of nonperturbative AdS/QCD models~\cite{Polchinski:2001tt}.  Although the mechanisms are different, both the pQCD and AdS/QCD approaches depend on the leading twist interpolating operators of the hadron and their structure at short distances. In both theories, hadronic form factors at high $Q^2$ are dominated by the wavefunctions at small impact separation. 
This in turn leads to the color transparency phenomena~\cite{Brodsky:1988xz,Bertsch:1981py}.
For example, measurements of pion photoproduction
are consistent with dimensional counting $s^{7}{ d\sigma/dt}(\gamma p \to \pi^+  n) \sim $ constant
at fixed CM angle for $s > 7 $ GeV.   The angular distributions seen in hard  large CM angle scattering reactions are consistent with quark interchange~\cite{Gunion:1973ex}, a result also predicted by the hard wall AdS/QCD model.
Reviews are given in Refs.~\cite{Sivers:1975dg} and~\cite{White:1994tj}.
One sees the onset of perturbative QCD scaling behavior even for
exclusive nuclear amplitudes such as deuteron photodisintegration
(Here $n = 1+ 6 + 3 + 3 = 13 $)
and
$s^{11}{ d\sigma/dt}(\gamma d \to p n) \sim $ constant at fixed CM
angle~\cite{Holt:1990ze,Bochna:1998ca,Rossi:2004qm}. The measured
deuteron form factor~\cite{Rock:1991jy} also appears to follow the
leading-twist QCD predictions~\cite{Brodsky:1976rz} at large
momentum transfers in the few GeV region.  The six color-triplet quarks of the valence Fock state of the deuteron can be arranged as a sum of five different color-singlet states, only one of which can be identified with the neutron-proton state and can account for the large magnitude of the deuteron form factor at high scales.
A measurement of ${d\sigma/dt}(\gamma d \to \Delta^{++}\Delta)$  in the scaling region
can establish the role of ``hidden-color" degrees of
freedom~\cite{Brodsky:1983vf} of the nuclear wavefunction in hard
deuteron reactions.

In the case of pQCD, the near-constancy of the
effective QCD coupling at small scales helps explain the general empirical
success of the dimensional counting rules for the near-conformal power
law fall-off of form factors and fixed-angle
scaling~\cite{Brodsky:1989pv}.

Color transparency~\cite{Brodsky:1988xz,Bertsch:1981py} is a key
property of color gauge theory, and it thus stands at the
foundations of QCD. Color transparency has been confirmed in
diffractive dijet production~\cite{Aitala:2000hc}, pion
photoproduction~\cite{Clasie:2007gq} and vector meson
electroproduction~\cite{Airapetian:2002eh}, but it is very important
to also systematically validate it in large-angle hadron scattering
processes.   Color transparency and higher-twist subprocesses
~\cite{Berger:1979du, Berger:1980qg, Berger:1981fr, Arleo:2010yg, Arleo:2009ch} where the trigger hadron is produced directly, such as $u u \to p \bar d,$ can account for the anomalous growth of the baryon-to-meson ratio with increasing centrality observed  in heavy ion collisions at RHIC~\cite{Brodsky:2008qp}.

\section{ Anomalies in Exclusive Processes}

Some exceptions to the general success of dimensional counting are known:

The transition form factor $F(Q^2)_{\gamma \to \pi^0}$ between a real photon and a pion has been recently measured at BaBar to high $Q^2 \simeq 10~{\rm GeV}^2$, falling at high photon virtuality roughly as $1/ Q$ rather than the predicted $1/Q^2$ fall-off.  In contrast, preliminary measurements from BaBar~\cite{Druzhinin:2009gq} indicate that the transition form factors $F(Q^2)_{\gamma \to \eta}$
and $F(Q^2)_{\gamma \to \eta^\prime}$ are consistent  with the pQCD expectations. The photon to meson transition form factor is the simplest QCD hadronic exclusive amplitude, and thus it is critical to understand this discrepancy. As we shall discuss below, AdS/QCD predicts a broad distribution amplitude $\phi_\pi(x,Q) $ in the nonperturbative domain, but since ERBL evolution leads to a narrower distribution in the high $Q$ domain,  it cannot explain the BaBar anomaly.  It is hard to imagine that the pion distribution amplitude is very flat~\cite{Radyushkin:2009zg, Polyakov:2009je, Mikhailov:2009sa} since this corresponds to a pointlike non-composite hadron.  It is  crucial to measure ${d\sigma\over dt }( \gamma \gamma \to  \pi^0 \pi^0)$  since the CM angular distribution is very sensitive to the shape of $\phi_\pi(x,Q)$~\cite{Brodsky:1981rp}.

The Hall A collaboration~\cite{Danagoulian:2007gs} at JLab
has reported another significant exception to the general empirical
success of dimensional counting in fixed-CM-angle Compton scattering
${d\sigma/dt}(\gamma p \to \gamma p) \sim {F(\theta_{CM})/s^8} $
instead of the predicted $1/s^6$ scaling.  The deviations from
fixed-angle conformal scaling may be due to corrections from
resonance contributions in the JLab energy range. It is interesting
that the hadron form factor $R_V(t)$~\cite{Diehl:1998kh}, which
multiplies the $\gamma q \to \gamma q$ amplitude is found by Hall A
to scale as $1/t^2$, in agreement with the  pQCD and AdS/QCD
prediction. In addition, the Belle measurement~\cite{Chen:2001sm} of
the timelike two-photon cross section ${d \sigma/dt}(\gamma \gamma
\to p \bar p)$ is consistent with $1/ s^6$ scaling.

Although large-angle proton-proton elastic scattering is well
described by dimensional scaling $s^{10}{ d\sigma/dt}(p p \to p p)
\sim $ constant at fixed CM angle,  extraordinarily  large spin-spin
correlations are observed~\cite{Court:1986dh}. The ratio of
scattering cross sections for spin-parallel and normal to the
scattering plane versus spin-antiparallel reaches $R_{NN} \simeq 4$
in large angle $ p p \to p p$ at $\sqrt s \simeq 5~$GeV; this is a
remarkable example of ``exclusive transversity".  Color transparency
is observed at lower energies but it fails~\cite{Mardor:1998zf} at
the same energy where $R_{NN}$ becomes large.  In fact, these
anomalies have a natural explanation~\cite{Brodsky:1987xw} as a
resonance effect related to the charm threshold in $pp$ scattering.
Alternative explanations of the large spin correlation are discussed
and reviewed in Ref.~\cite{Dutta:2004fw}.
Resonance formation is a natural phenomenon when all constituents
are relatively at rest. For example, a resonance effect can occur
due to the intermediate state  $uud uud c \bar c$ at the charm
threshold $\sqrt s = 5$ GeV in $p p$ collisions. Since the $c $ and
$\bar c$ have opposite intrinsic parity, the resonance appears in
the $L= J= S =1$ partial wave for $ p p \to p p$ which is only
allowed for spin-parallel and normal
to the  scattering plane
$A_{NN}=1$~\cite{Brodsky:1987xw}. Resonance formation at the charm
threshold also explains the dramatic quenching of color transparency
seen in quasielastic $p n$ scattering by the EVA BNL
experiment~\cite{Mardor:1998zf} in the same kinematic region. The
reason why these effects are apparent in $p p \to p p$ scattering
is that the amplitude for the formation of an $uud uud c \bar c$
$s$-channel resonance in the intermediate state is of the same
magnitude as the fast-falling background $p p \to  p p$ pQCD
amplitude from quark interchange at large CM angles: $M(p p \to pp)
\sim {1/u^2 t^2} .$   The open charm cross
section in $p p$ scattering  is predicted by unitarity to be of order of $1 ~\mu b$ at
threshold~\cite{Brodsky:1987xw}.  One also expects similar novel QCD phenomena in
large-angle photoproduction $\gamma p \to \pi N$ near the charm
threshold, including the breakdown of color transparency and strong
spin-spin correlations. These effects  can be tested by measurements
at the new JLab 12 GeV facility,  which would confirm resonance
formation in a low partial wave in $\gamma p \to \pi N$ at $\sqrt s
\simeq 4$ GeV due to attractive forces in the $uu d \bar c c$
channel.

Another difficulty for the application of pQCD to exclusive processes is the famous $J/\psi \to \rho \pi$ puzzle;  the observed unusually large branching ratio for $J/\psi \to \rho \pi$. In contrast, the branching ratio for $\Psi^\prime \to \rho \pi$ is very small. Such decays into pseudoscalar plus vector mesons require light-quark helicity suppression or internal orbital angular momentum and thus should be suppressed by hadron helicity conservation in pQCD.
However, the $J/\psi \to \rho \pi$  puzzle can be explained by the presence of intrinsic charm Fock states in the outgoing mesons~\cite{Brodsky:1997fj}.

\section{Light-Front Quantization and Exclusive Processes}

Light-front (LF) quantization is the ideal framework for  describing the
structure of hadrons in terms of their quark and gluon degrees of
freedom. The
light-front wavefunctions
(LFWFs) of bound states in QCD are relativistic generalizations of the Schr\"odinger wavefunctions, but they are determined at fixed light-front time $\tau = x^+ = x^0 + x^3$, the time marked by the
front of a light wave~\cite{Dirac:1949cp}, rather than at fixed ordinary time
$t.$ They play the same role in
hadron physics that Schr\"odinger wavefunctions play in atomic physics.
In addition, the simple structure of the LF vacuum provides an unambiguous
definition of the partonic content of a hadron in QCD.
In the light-front formalism, one  sets boundary conditions at fixed $\tau$ and then evolves the system using the LF Hamiltonian
$P^-  \! =  \! P^0 \! - P^3 = i {d/d \tau}$.  The invariant Hamiltonian $H_{LF}  \! =  P_\mu P^\mu = P^+ P^- \! - P^2_\perp$
has eigenvalues $\mathcal{M}^2$ where $\mathcal{M}$
is the physical hadron mass.
The Heisenberg equation for QCD on the light-front thus takes the form $H_{LF} \vert \Psi_H \rangle = \mathcal{M}^2_H \vert \Psi_H \rangle $,  where $H_{LF}$ is determined canonically from the QCD Lagrangian. Its eigenfunctions are the light-front eigenstates which define the frame-independent light-front wavefunctions, and
its eigenvalues yield the hadronic spectrum, the bound states as well as the continuum.  The  projection of the eigensolutions on the free Fock basis gives the $n$-parton LF wavefunctions
$\psi_{n/H} = \langle n \vert \Psi_H \rangle$ needed for phenomenology.   Heisenberg's problem on the light-front can be solved numerically using discretized light-front quantization (DLCQ)~\cite{Pauli:1985ps} by applying anti-periodic boundary conditions in $\sigma = x^0-x^3.$  This method has been used successfully to solve many lower dimension quantum field theories~\cite{Brodsky:1997de}.

The light-front Fock state wavefunctions
$\psi_{n/H}(x_i, \vec k_\perp, \lambda_i)$ are functions of LF momentum fractions $x_i = {k^+_i\over P^+} = {k^0_i+k^3_i\over P^0+P^3}$ with $\sum^n_{i=1} x_i =1,$  relative transverse momenta satisfying 
$\sum^n_{i=1} \vec k_{\perp i} = 0$,
 and spin projections $\lambda_i.$  Remarkably, the LFWFs are independent of the hadron's total momentum $P^+ = P^0 \! + P^3$, so that once they are known in one frame, they are known in all frames; Wigner transformations and Melosh rotations
are not required. The light-front formalism for gauge theories in
light-cone gauge is particularly useful in that there are no ghosts
and one has a direct physical interpretation of  orbital angular
momentum.  They also allow one to formulate hadronization in inclusive and exclusive reactions at the amplitude level.

A key example of the utility of the light-front is the Drell-Yan West formula~\cite{Drell:1969km, West:1970av}}
for the spacelike form factors of electromagnetic currents given as overlaps of initial and final LFWFs. At high momentum where one can iterate the hard scattering kernel, this yields the dimensional counting rules, factorization theorems, and ERBL evolution of the distribution amplitudes. The gauge-invariant distribution
amplitudes $\phi_H(x_i,Q)$ defined from the integral over the
transverse momenta $\mbf{k}^2_{\perp i} \le Q^2$ of the valence
(smallest $n$) Fock state provide a fundamental measure of the
hadron at the amplitude level~\cite{Lepage:1979zb, Efremov:1979qk};
they  are the nonperturbative inputs to the factorized form of hard
exclusive amplitudes and exclusive heavy hadron decays in
pQCD.

Given the light-front wavefunctions $\psi_{n/H}$ one can
compute a large range of other hadron
observables. For example, the valence and sea quark and gluon
distributions which are measured in deep inelastic lepton scattering
are defined from the squares of the LFWFs summed over all Fock
states $n$. Exclusive weak transition
amplitudes~\cite{Brodsky:1998hn} such as $B\to \ell \nu \pi$,  and
the generalized parton distributions~\cite{Brodsky:2000xy} measured
in deeply virtual Compton scattering  $\gamma^* p \to \gamma p$ are (assuming the ``handbag"
approximation) overlaps of the initial and final LFWFs with $n
=n^\prime$ and $n =n^\prime+2$. The resulting distributions obey the DGLAP and
ERBL evolution equations as a function of the maximal invariant
mass, thus providing a physical factorization
scheme~\cite{Lepage:1980fj}. In each case, the derived quantities
satisfy the appropriate operator product expansions, sum rules, and
evolution equations. At large $x$ where the struck quark is
far-off shell, DGLAP evolution is quenched~\cite{Brodsky:1979qm}, so
that the fall-off of the DIS cross sections in $Q^2$ satisfies Bloom-Gilman
inclusive-exclusive duality at fixed $W^2.$

The simple features of the light-front contrast with the conventional instant form where one quantizes at $t=0.$ For example, calculating a hadronic form factor requires boosting the hadron's wavefunction from the initial to final state, a dynamical problem  as difficult as solving QCD itself. Moreover current matrix elements require computing the interaction of the probe with all of connected currents fluctuating in the QCD vacuum.  Each contributing diagram is frame-dependent.

A fundamental theorem for gravity can be derived from the equivalence principle:  the anomalous gravitomagnetic moment defined from the spin-flip  matrix element of the energy-momentum tensor is identically zero $B(0)=0$~\cite{Teryaev:1999su}. This theorem can be proven in  the light-front formalism Fock state by Fock state~\cite{Brodsky:2000ii}. The LF vacuum is trivial up to zero modes in the front form, thus eliminating contributions to the cosmological constant from QED or QCD~\cite{Brodsky:2009zd}.

\section{AdS/QCD}
One of the most significant theoretical advances in recent years has
been the application of the AdS/CFT
correspondence~\cite{Maldacena:1997re} between string theories
defined in 5-dimensional Anti--de Sitter (AdS) space-time and
conformal field theories in physical space-time,
to study  the dynamics of strongly coupled quantum field theories.
The essential principle
underlying the AdS/CFT approach to conformal gauge theories is the
isomorphism of the group of Poincar\'{e} and conformal transformations
$SO(4,2)$ to the group of isometries of Anti-de Sitter space.  The
AdS metric is
\begin{equation} \label{eq:AdSz}
ds^2 = \frac{R^2}{z^2}(\eta_{\mu \nu} dx^\mu
 dx^\nu - dz^2),
 \end{equation}
which is invariant under scale changes of the
coordinate in the fifth dimension $z \to \lambda z$ and $ x_\mu \to
\lambda x_\mu$.  Thus one can match scale transformations of the
theory in $3+1$ physical space-time to scale transformations in the
fifth dimension $z$.
In the AdS/CFT duality, the amplitude $\Phi(z)$ represents the
extension of the hadron into the additional fifth dimension.  The
behavior of
$\Phi(z) \to z^\tau$ at $z \to 0$
matches the twist-dimension $\tau$ of the hadron at short distances.

QCD is not conformal but  there is in fact much empirical evidence from lattice gauge theory~\cite{Furui:2006py}, Dyson Schwinger theory~\cite{vonSmekal:1997is},  and empirical effective charges~\cite{Deur:2005cf}, that the QCD $\beta$-function vanishes in the infrared~\cite{Deur:2008rf}.  The QCD infrared fixed point arises since the propagators of the confined quarks and gluons in the  loop integrals contributing to the $\beta$-function have a maximal wavelength~\cite{Brodsky:2008be}. The decoupling of quantum loops in the infrared is analogous to QED where vacuum polarization corrections to the photon propagator decouple at $Q^2 \to 0$.
Since there is a  window where the QCD coupling is
large and approximately constant,
QCD resembles a conformal theory for massless quarks. Thus, even though QCD is not conformally invariant,
one can use the mathematical representation of the conformal group
in five-dimensional Anti-de Sitter space to construct an analytic
first approximation to the theory.

The AdS/QCD correspondence is now providing important insight into the properties of QCD needed to compute exclusive reactions. In particular, the soft-wall AdS/QCD model modified by a positive sign dilaton metric, which represents color confinement, leads to a remarkable one-parameter description of nonperturbative hadron dynamics, including successful predictions for the meson and baryon spectrum for zero quark mass,
including
a zero-mass pion, a Regge spectrum of linear trajectories with the same slope in orbital angular $L$ and the
principal quantum number $n$, as well as dynamical form factors.
The theory predicts
dimensional counting for form factors and other fixed CM angle exclusive reactions.  Moreover, as we shall review,
light-front holography allows one to map the hadronic amplitudes $\phi_H(z)$ determined in the AdS fifth dimension $z$ to the valence LFWFs of each hadron as a function of a covariant impact variable $\zeta.$  Moreover, the same techniques provide a prediction for the QCD coupling $\alpha_s(Q^2)$ and its $\beta$-function which reflects the dynamics of confinement.

\section{AdS/QCD Models}

We thus begin with a conformal approximation to QCD to model an effective dual gravity description in AdS space. The  five-dimensional AdS$_5$ geometrical representation of the conformal group represents scale transformations within the conformal window.  Confinement can be
 introduced with a sharp cut-off in the infrared region of AdS space, as in the ``hard-wall" model~\cite{Polchinski:2001tt},
 or, more successfully,  using a dilaton background in the fifth dimension to produce a smooth cutoff at large distances
as  in the ``soft-wall" model~\cite{Karch:2006pv}.
 We assume a dilaton profile 
 $\exp(+\kappa^2 z^2)$~\cite{deTeramond:2009xk,Andreev:2006ct,Zuo:2009dz,Afonin:2010fr},
 with  opposite sign  to that of Ref.~\cite{Karch:2006pv}.
 The soft-wall AdS/QCD model
 with a positive-sign dilaton-modified AdS metric,~\cite{Andreev:2006ct}
$ds^2 = (R^2/z^2)
 e^{+{\kappa^2 z^2}}(\eta_{\mu \nu} dx^\mu
 dx^\nu - dz^2),$
leads to the harmonic potential~\cite{deTeramond:2009xk} 
$U(z) = \kappa^4 z^2
+ 2 \kappa^2(L+S-1)$~in the fifth dimension coordinate $z$.
the resulting spectrum reproduces linear Regge trajectories.
${\cal M}^2(S,L,n) = 4 \kappa^2(n + L + S/2)$,
where ${\cal M}^2(S,L,n) $ is proportional to the internal
spin,  orbital angular momentum $L$ and the
principal radial
quantum number $n$.

The modified metric induced by the dilaton can be interpreted in AdS space as a gravitational potential
for an object of mass $m$  in the fifth dimension:
$V(z) = mc^2 \sqrt{g_{00}} = mc^2 R \, e^{\pm \kappa^2 z^2/2}/z$.
In the case of the negative solution, the potential decreases monotonically, and thus an object in AdS will fall to infinitely large
values of $z$.  For the positive solution, the potential is non-monotonic and has an absolute minimum at $z_0 = 1/\kappa$.
Furthermore, for large values of $z$ the gravitational potential increases exponentially,  confining any object  to distances
$\langle z \rangle \sim 1/\kappa$~\cite {deTeramond:2009xk}.  We thus will choose the confining positive sign dilaton solution~\cite{deTeramond:2009xk,Andreev:2006ct} with  opposite sign  to that of 
Ref.~\cite{Karch:2006pv}. This additional warp factor leads to a well-defined scale-dependent effective coupling.

Glazek and Schaden~\cite{Glazek:1987ic} have shown that a  harmonic oscillator confining potential naturally arises as an effective potential between heavy quark states when one stochastically eliminates higher gluonic Fock states. Also, Hoyer~\cite{Hoyer:2009ep} has argued that the Coulomb  and   linear  potentials are uniquely allowed in the Dirac equation at the classical level. The linear potential  becomes a harmonic oscillator potential in the corresponding Klein-Gordon equation.

\section{Light-Front Holography}

Light-front  holography \cite{deTeramond:2008ht, Brodsky:2006uqa, Brodsky:2007hb, Brodsky:2008pf,  deTeramond:2010we} connects
the equations of motion in AdS space and
the Hamiltonian formulation of QCD in physical space-time quantized
on the light front  at fixed LF time.  This correspondence provides a direct connection between the hadronic amplitudes $\Phi(z)$  in AdS space  with  LF wavefunctions $\phi(\zeta)$ describing the quark and gluon constituent structure of hadrons in physical space-time.
In the case of a meson, $\zeta = \sqrt{x(1-x) {\bf b}^2_\perp}$ is a Lorentz invariant coordinate which measures
the distance between the quark and antiquark; it is analogous to the radial coordinate $r$ in the Schr\"odinger equation. Here $\vec b_\perp$ is the Fourier conjugate of the transverse
momentum $\vec k_\perp$.
In effect $\zeta$ represents the off-light-front energy shell and invariant mass dependence of the bound state; it allows the separation of the dynamics of quark and gluon binding from the kinematics of constituent spin and internal orbital angular momentum~\cite{deTeramond:2008ht}.
Light-front holography thus provides a  connection between the description of
hadronic modes in AdS space and the Hamiltonian formulation of QCD in
physical space-time quantized on the light-front  at fixed LF
time $\tau.$ The resulting equation for the mesonic $q \bar q$ bound states at fixed light-front time  has the form of a single-variable relativistic Lorentz invariant
 Schr\"odinger equation~\cite{deTeramond:2008ht}
\begin{equation} \label{eq:QCDLFWE}
\left(-\frac{d^2}{d\zeta^2}
- \frac{1 - 4L^2}{4\zeta^2} + U(\zeta) \right)
\phi(\zeta) = \mathcal{M}^2 \phi(\zeta),
\end{equation}
where the confining potential is $ U(\zeta) = \kappa^4 \zeta^2 + 2 \kappa^2(L+S-1)$
in the soft-wall model
with a positive-sign dilaton-modified AdS metric~\cite{deTeramond:2009xk}.
Its eigenvalues determine the hadronic spectra and its eigenfunctions are related to the light-front wavefunctions
of hadrons for general spin and orbital angular momentum. This
LF wave equation serves as a semiclassical first approximation to QCD,
and it is equivalent to the equations of motion which describe the
propagation of spin-$J$ modes in  AdS space.  The resulting light-front wavefunctions provide a fundamental description of the structure and internal dynamics of hadronic states in terms of their constituent quark and gluons.
There is only one parameter, the mass scale $\kappa \sim 1/2$ GeV, which enters the confinement potential. In the case of mesons $S=0,1$ is the combined spin of the $q $ and $ \bar q $ state, $L$ is their relative orbital angular momentum as determined by the hadronic light-front wavefunctions.

The mapping between the LF invariant variable $\zeta$ and the fifth-dimension AdS coordinate $z$ was originally obtained
by matching the expression for electromagnetic current matrix
elements in AdS space  with the corresponding expression for the
current matrix element, using LF  theory in physical space
time~\cite{Brodsky:2006uqa}.   It has also been shown that one
obtains the identical holographic mapping using the matrix elements
of the energy-momentum tensor~\cite{Brodsky:2008pf,Abidin:2008ku}, thus  verifying  the  consistency of the holographic
mapping from AdS to physical observables defined on the light front.

\section{The Hadron Spectrum
and Form Factors
in Light-Front AdS/QCD}

The meson spectrum predicted by  Eq. \ref{eq:QCDLFWE} has a string-theory Regge form
${\cal M}^2 = 4 \kappa^2(n+ L+S/2)$; {\it i.e.}, the square of the eigenmasses are linear in both the orbital angular momentum $L$ and $n$, where $n$ counts the number of nodes  of the wavefunction in the radial variable $\zeta$.
The spectrum also depends on the internal spin S.
This is illustrated for the pseudoscalar and vector meson spectra in Fig. \ref{pionVM},
where the data are from Ref.~\cite{Amsler:2008xx}.
The pion ($S=0, n=0, L=0$) is massless for zero quark mass, consistent with the chiral invariance of massless
quarks in
QCD.  Thus one can compute the hadron spectrum by simply adding  $4 \kappa^2$ for a unit change in the radial quantum number, $4 \kappa^2$ for a change in one unit in the orbital quantum number  $L$ and $2 \kappa^2$ for a change of one unit of spin $S$. Remarkably, the same rule holds for three-quark baryons as we shall show below.

\begin{figure}[h]
\begin{center}
\includegraphics[width=6.8cm]{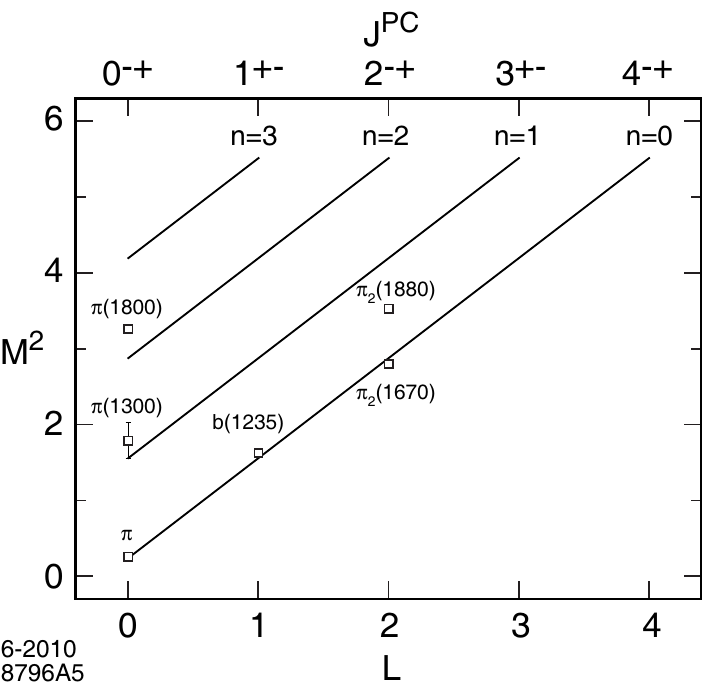}  ~~~
\includegraphics[width=6.8cm]{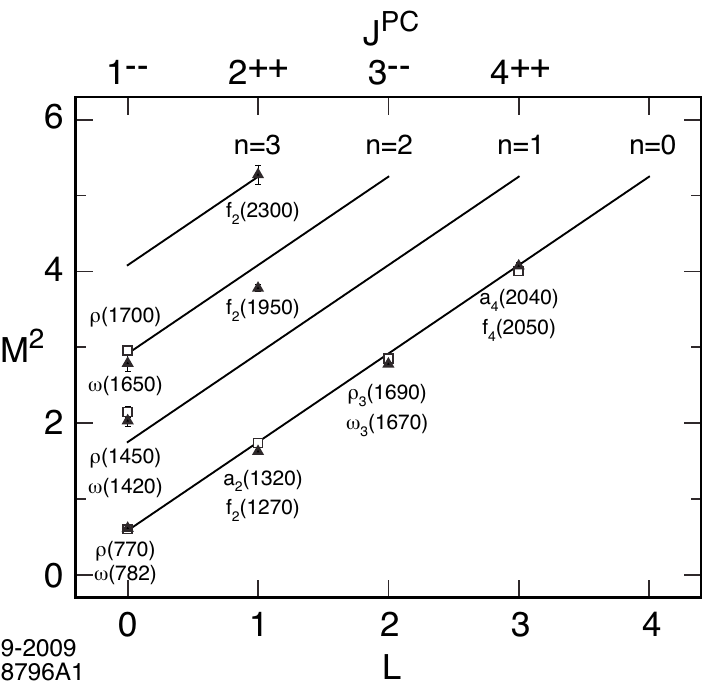}
 \caption{Parent and daughter Regge trajectories for (a) the $\pi$-meson family with
$\kappa= 0.6$ GeV; and (b) the  $I\!=\!1$ $\rho$-meson
 and $I\!=\!0$  $\omega$-meson families with $\kappa= 0.54$ GeV.}
\label{pionVM}
\end{center}
\end{figure}

The eigensolutions of  Eq. \ref{eq:QCDLFWE} provide the light-front wavefunctions of the valence Fock state of the hadrons $\psi(x,
\vec b_\perp)$  as illustrated for the pion in Fig. \ref{LFWFPionFFSL} for the soft-wall (a) and hard-wall (b) models.   The resulting distribution amplitude has a
broad form $\phi_\pi(x) \sim \sqrt{x(1-x)}$ which is compatible with moments determined from lattice gauge theory. One can then immediately
compute observables such as hadronic form factors (overlaps of LFWFs), structure functions (squares of LFWFs), as well as the generalized parton
distributions and distribution amplitudes which underly hard exclusive reactions. For example, hadronic form factors can be predicted from the
overlap of LFWFs in the Drell-Yan West formula. The prediction for the space-like pion form factor is shown in Fig.  \ref{LFWFPionFFSL} (c). The
pion form factor and the vector meson poles residing in the dressed current in the soft wall model require choosing  a value of $\kappa$ smaller
by a factor of $1/\sqrt 2$  than the canonical value of  $\kappa$ which determines the mass scale of the hadronic spectra.  This shift is
apparently due to the fact that the transverse current in $e^+ e^- \to q \bar q$ creates a quark pair with $L^z= \pm 1$ instead of the $L^z=0$
$q \bar q$ composition of the vector mesons in the spectrum.

\begin{figure}[h]
\begin{center}
 \includegraphics[width=8.0cm]{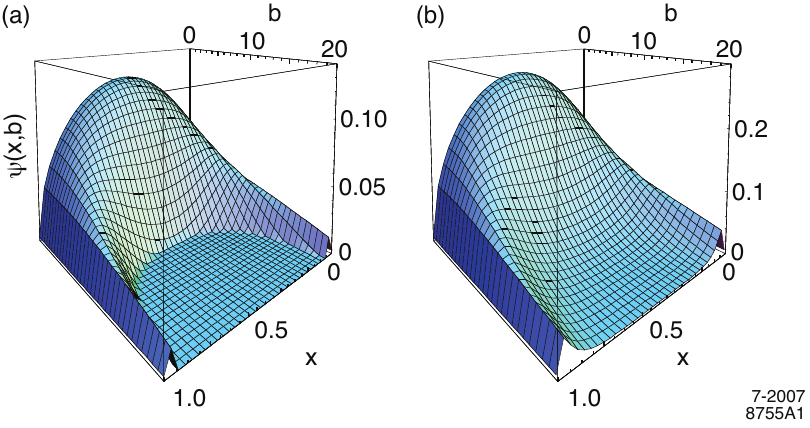}~~
\includegraphics[width=5.0cm]{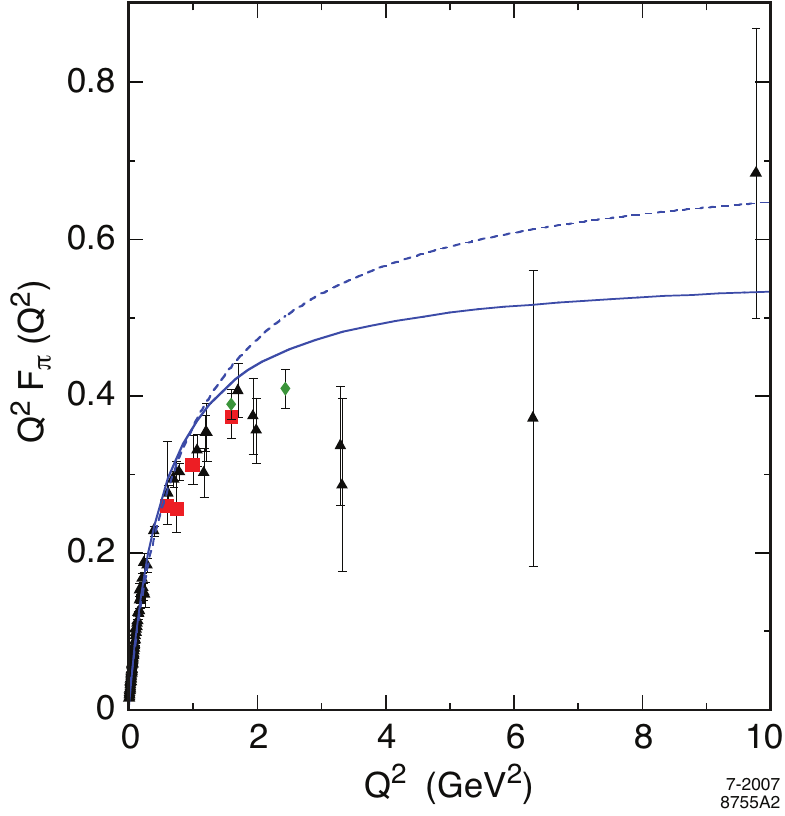}
 \caption{ Pion LF wavefunction $\psi_\pi(x, \vec{b}_\perp$) for the  AdS/QCD (a) hard-wall ($\Lambda_{QCD} = 0.32$ GeV) and (b) soft-wall  ($\kappa = 0.375$ GeV)  models.  (c) Space-like scaling behavior for $Q^2 F_\pi(Q^2).$ The continuous line is the prediction of the soft-wall model for
$\kappa = 0.375$ GeV.  The dashed line is the prediction of the hard-wall model for $\Lambda_{\rm QCD} = 0.22$ GeV. The triangles are the data
compilation of Baldini
{\it et al.}~\cite{Baldini:1998qn},  the  boxes  are JLAB 1 data~\cite{Tadevosyan:2007yd} and  the diamonds are JLAB 2
data~\cite{Horn:2006tm}.}
\label{LFWFPionFFSL}
\end{center}
\end{figure}

Individual hadrons in AdS/QCD are identified by matching the power behavior of the hadronic amplitude at the AdS boundary at small $z$ to the twist $\tau$ of its interpolating operator at short distances
as required by the AdS/CFT dictionary. The twist also equals the dimension of fields appearing in chiral super-multiplets~\cite{Craig:2009rk}; thus the twist of a hadron equals the number of constituents plus the relative orbital angular momentum.
One then can apply light-front holography to relate the amplitude eigensolutions  in the fifth-dimension coordinate $z$  to the LF wavefunctions in the physical space-time variable  $\zeta$.

Equation (\ref{eq:QCDLFWE}) was derived by taking the LF bound-state Hamiltonian equation of motion as the starting
point~\cite{deTeramond:2008ht}. The term $L^2/ \zeta^2$  in the  LF equation of motion  (\ref{eq:QCDLFWE})
is derived from  the reduction of the LF kinetic energy when one transforms
to two-dimensional cylindrical coordinates $(\zeta, \varphi)$,
in analogy to the $\ell(\ell+1)/ r^2$ Casimir term in Schr\"odinger theory.  One thus establishes the interpretation of $L$ in the AdS equations of motion.
The interaction terms build confinement  corresponding to
the dilaton modification of AdS space~\cite{deTeramond:2008ht}.
The duality between these two methods provides a direct
connection between the description of hadronic modes in AdS space and
the Hamiltonian formulation of QCD in physical space-time quantized
on the light-front  at fixed LF time $\tau.$

The identification of orbital angular momentum of the constituents is a key element in the description of the internal structure of hadrons using holographic principles. In our approach  quark and gluon degrees of freedom are explicitly introduced in the gauge/gravity correspondence~\cite{Brodsky:2003px}, in contrast with the usual
AdS/QCD framework~\cite{Erlich:2005qh,DaRold:2005zs} where axial and vector currents become the primary entities as in effective chiral theory.
Unlike the top-down string theory approach,  one is not limited to hadrons of maximum spin
$J \le 2$, and one can study baryons with finite color $N_C=3.$   Higher spin modes follow from shifting dimensions in the AdS wave equations.
In the soft-wall
model the usual Regge behavior is found $\mathcal{M}^2 \sim n +
L$, predicting the same multiplicity of states for mesons
and baryons as observed experimentally~\cite{Klempt:2007cp}.
It is possible to extend the model to hadrons with heavy quark constituents
by introducing nonzero quark masses and short-range Coulomb
corrections.  For other
recent calculations of the hadronic spectrum based on AdS/QCD, see Refs.~\cite{BoschiFilho:2005yh,Evans:2006ea, Hong:2006ta, Colangelo:2007pt, Forkel:2007ru, Vega:2008af, Nawa:2008xr, dePaula:2008fp,  Colangelo:2008us, Forkel:2008un, Ahn:2009px, Sui:2009xe, Kapusta:2010mf, Zhang:2010bn, Wang:2010aj}.
Other recent computations of the pion form factor are given in
Refs.~\cite{Kwee:2007dd,Grigoryan:2007wn, Bayona:2010bg}.

For baryons, the light-front wave equation is a linear equation
determined by the LF transformation properties of spin 1/2 states. A linear confining potential
$U(\zeta) \sim \kappa^2 \zeta$ in the LF Dirac
equation leads to linear Regge trajectories~\cite{Brodsky:2008pg}.   For fermionic modes the light-front matrix
Hamiltonian eigenvalue equation $D_{LF} \vert \psi \rangle = \mathcal{M} \vert \psi \rangle$, $H_{LF} = D_{LF}^2$,
in a $2 \times 2$ spinor  component
representation is equivalent to the system of coupled linear equations
\begin{eqnarray} \label{eq:LFDirac} \nonumber
- \frac{d}{d\zeta} \psi_- -\frac{\nu+{1\over 2}}{\zeta}\psi_-
- \kappa^2 \zeta \psi_-&=&
\mathcal{M} \psi_+, \\ \label{eq:cD2k}
  \frac{d}{d\zeta} \psi_+ -\frac{\nu+{1\over 2}}{\zeta}\psi_+
- \kappa^2 \zeta \psi_+ &=&
\mathcal{M} \psi_-,
\end{eqnarray}
with eigenfunctions
\begin{eqnarray} \nonumber
\psi_+(\zeta) &\sim& z^{\frac{1}{2} + \nu} e^{-\kappa^2 \zeta^2/2}
  L_n^\nu(\kappa^2 \zeta^2) ,\\
\psi_-(\zeta) &\sim&  z^{\frac{3}{2} + \nu} e^{-\kappa^2 \zeta^2/2}
 L_n^{\nu+1}(\kappa^2 \zeta^2),
\end{eqnarray}
and  eigenvalues $\mathcal{M}^2 = 4 \kappa^2 (n + \nu + 1)$.

\begin{figure}[h]
\begin{center}
\includegraphics[width=13.6cm]{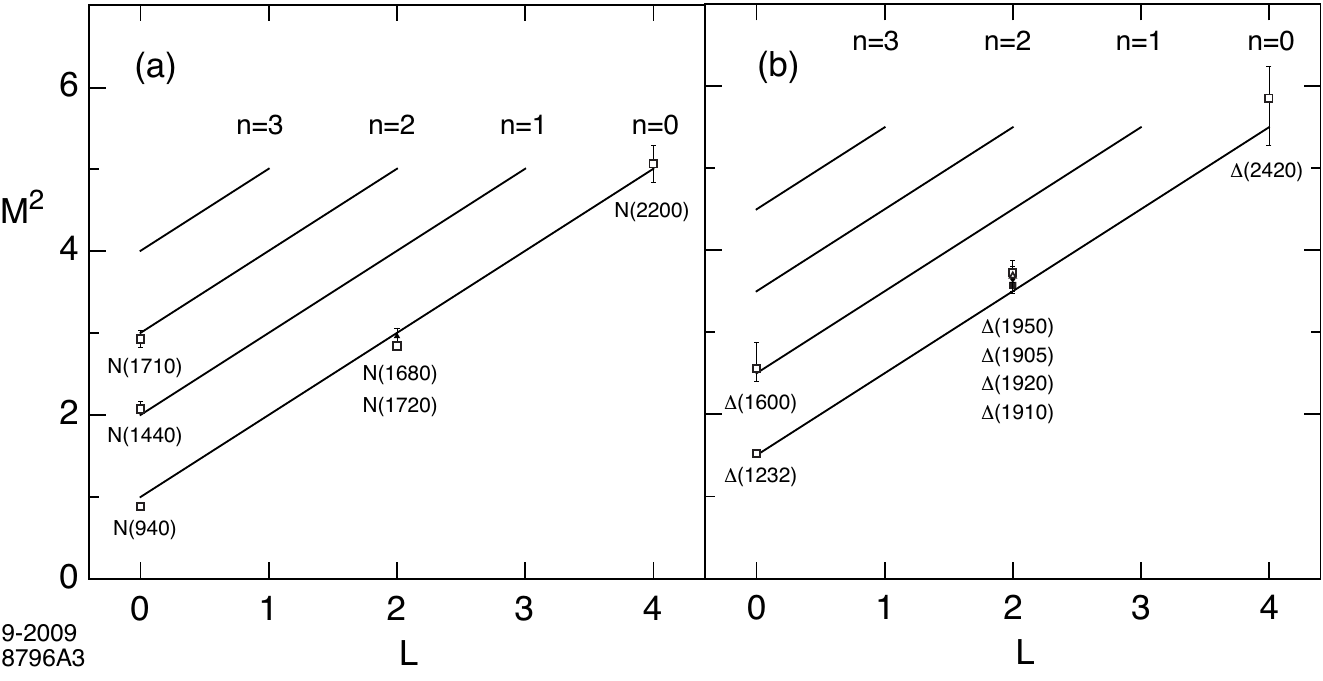}
\caption{{\bf 56}-plet Regge trajectories for  the  $N$ and $\Delta$ baryon families for $\kappa= 0.5$ GeV.}
\label{Baryons}
\end{center}
\end{figure}

The baryon interpolating operator
$ \mathcal{O}_{3 + L} =  \psi D_{\{\ell_1} \dots
 D_{\ell_q } \psi D_{\ell_{q+1}} \dots
 D_{\ell_m\}} \psi$,  $L = \sum_{i=1}^m \ell_i$, is a twist 3,  dimension $9/2 + L$ with scaling behavior given by its
 twist-dimension $3 + L$. We thus require $\nu = L+1$ to match the short distance scaling behavior. Higher spin modes are obtained by shifting dimensions for the fields.
 Thus, as in the meson sector,  the increase  in the
mass squared for higher baryonic state is
$\Delta n = 4 \kappa^2$, $\Delta L = 4 \kappa^2$ and $\Delta S = 2 \kappa^2,$
relative to the lowest ground state,  the proton. Since our starting point to find the bound state equation of motion for baryons is the light-front, we fix the overall energy scale identical for mesons and baryons by imposing chiral symmetry to the pion~\cite{deTeramond:2010we} in the LF Hamiltonian equations. By contrast, if we start with a five-dimensional action for a scalar field in presence of a positive sign dilaton, the pion is automatically massless.

The predictions for the $\bf 56$-plet of light baryons under the $SU(6)$  flavor group are shown in Fig. \ref{Baryons}.
As for the predictions for mesons in Fig. \ref{pionVM}, only confirmed PDG~\cite{Amsler:2008xx} states are shown.
The Roper state $N(1440)$ and the $N(1710)$ are well accounted for in this model as the first  and second radial
states. Likewise the $\Delta(1660)$ corresponds to the first radial state of the $\Delta$ family. The model is  successful in explaining the important parity degeneracy observed in the light baryon spectrum, such as the $L\! =\!2$, $N(1680)\!-\!N(1720)$ degenerate pair and the $L=2$, $\Delta(1905), \Delta(1910), \Delta(1920), \Delta(1950)$ states which are degenerate
within error bars. Parity degeneracy of baryons is also a property of the hard wall model, but radial states are not well described in this model~\cite{deTeramond:2005su}.

As an example  of the scaling behavior of a twist $\tau = 3$ hadron, we compute the spin non-flip
nucleon form factor in the soft wall model~\cite{Brodsky:2008pg}. The proton and neutron Dirac
form factors are given by
\begin{equation}
F_1^p(Q^2) =  \! \int  d \zeta \, J(Q, \zeta) \,
  \vert \psi_+(\zeta)\vert^2 ,
\end{equation}
\begin{equation}
F_1^n(Q^2) =  - \frac{1}{3}  \! \int  d \zeta  \,  J(Q, \zeta)
 \left[\vert \psi_+(\zeta)\vert^2 - \vert\psi_-(\zeta)\vert^2\right],
 \end{equation}
where $F_1^p(0) = 1$,~ $F_1^n(0) = 0$. The non-normalizable mode
 $J(Q,z)$ is the solution of the
AdS wave equation for the external electromagnetic current in presence of a dilaton
background  $\exp(\pm \kappa^2 z^2)$~\cite{Brodsky:2007hb, Grigoryan:2007my}.
Plus and minus components of the twist 3 nucleon LFWF are
\begin{equation} \label{eq:PhipiSW}
\psi_+(\zeta) \!=\! \sqrt{2} \kappa^2 \, \zeta^{3/2}  e^{-\kappa^2 \zeta^2/2},  ~~~
\Psi_-(\zeta) \!=\!  \kappa^3 \, \zeta^{5/2}  e^{-\kappa^2 \zeta^2/2}.
\end{equation}
The results for $Q^4 F_1^p(Q^2)$ and $Q^4 F_1^n(Q^2)$   and are shown in
Fig. \ref{fig:nucleonFF}

\begin{figure}[h]
\begin{center}
 \includegraphics[width=7.6cm]{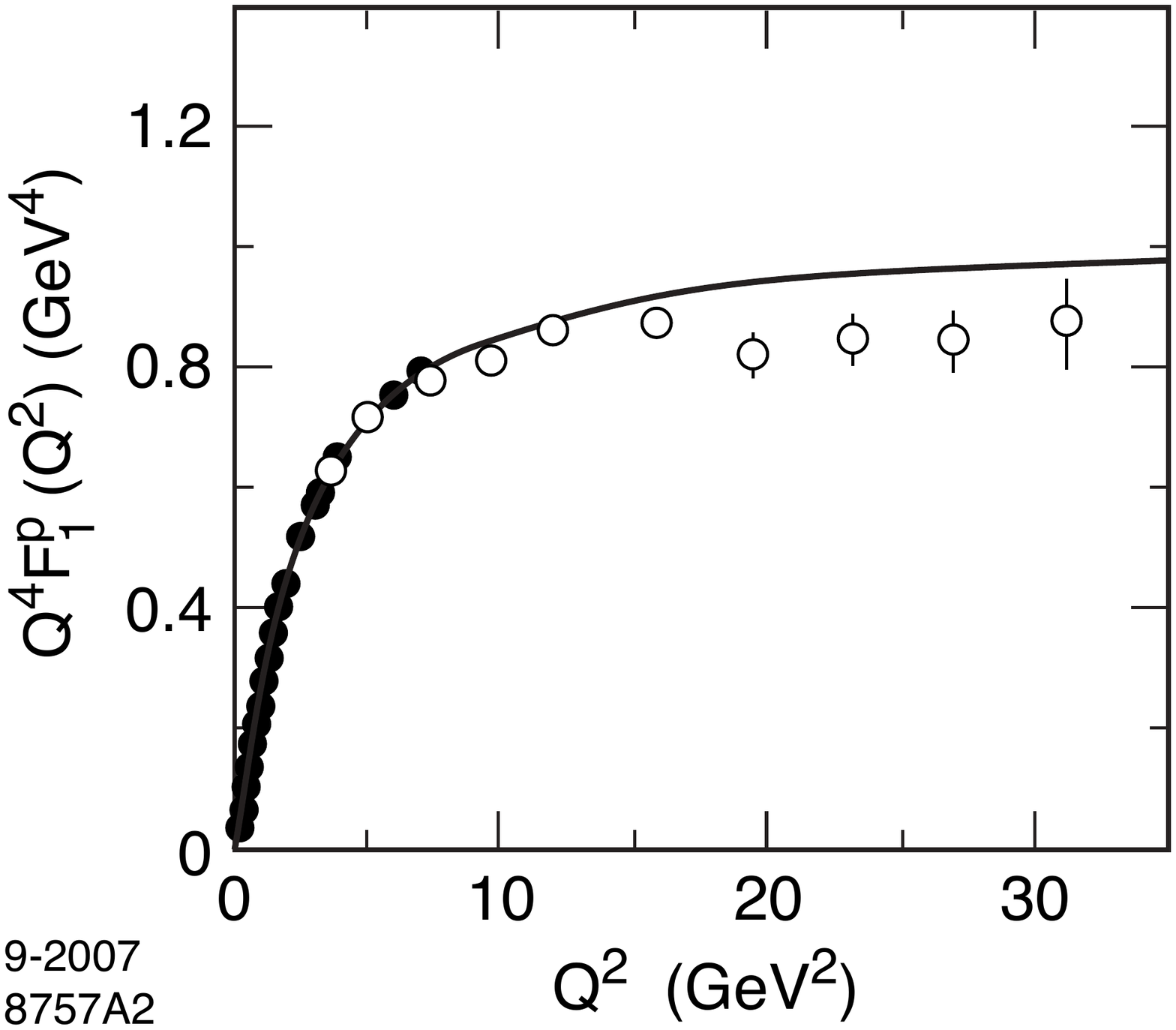}
\includegraphics[width=7.4cm]{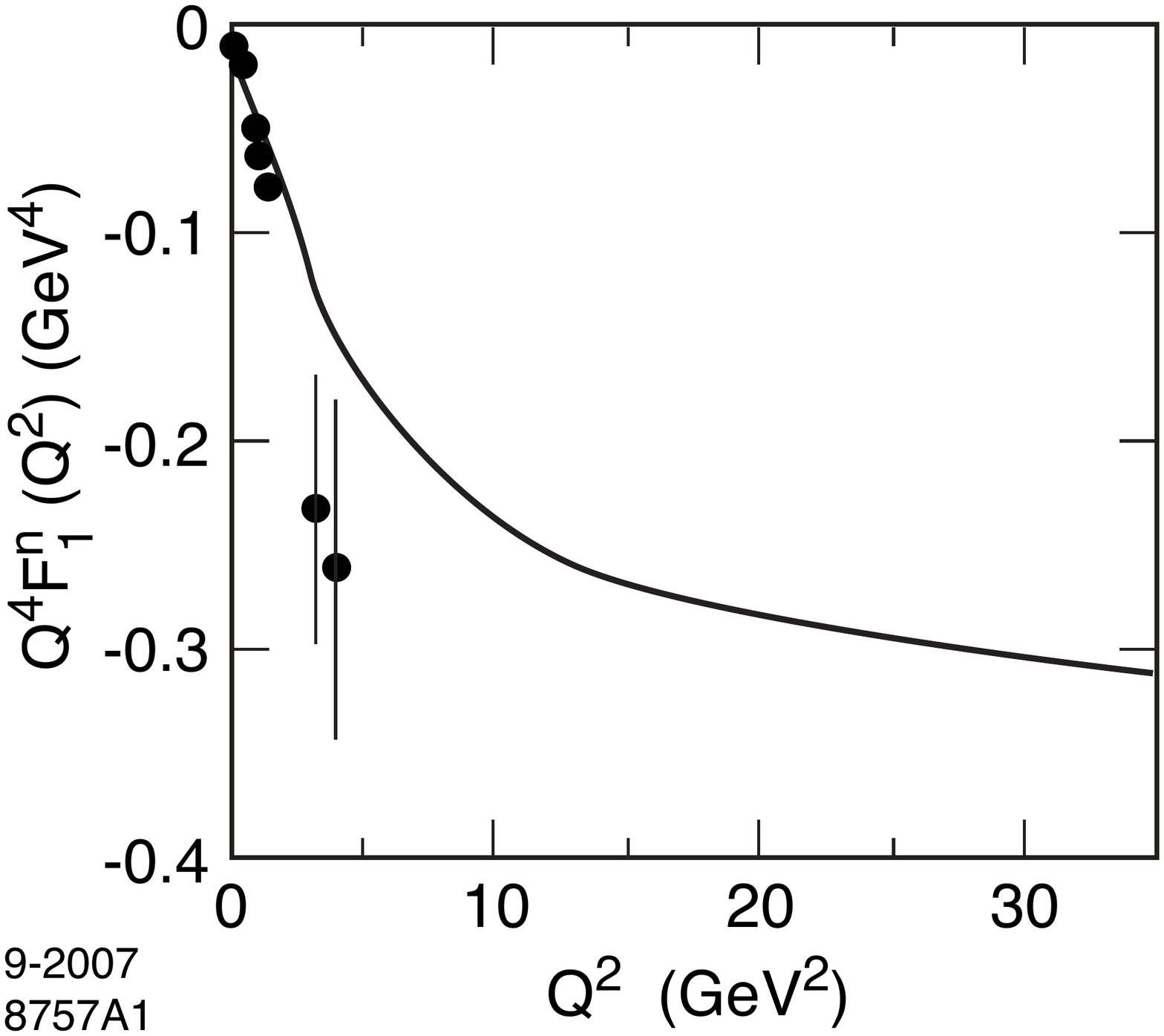}
 \caption{Predictions for $Q^4 F_1^p(Q^2)$ and $Q^4 F_1^n(Q^2)$ in the
soft wall model for $\kappa =  0.424$ GeV~\cite{note3}.}
\label{fig:nucleonFF}
\end{center}
\end{figure}

\section{Nonperturbative Running Coupling from Light-Front Holography \label{alphaAdS}}

The concept of a running coupling $\alpha_s(Q^2)$  in QCD is usually restricted to the perturbative domain.  However, as in QED, it is useful to define the coupling as an analytic function valid over the full space-like and time-like domains.
The study of the non-Abelian QCD coupling at small momentum transfer is a complex problem because of  gluonic self-coupling and color confinement.

The definition of the running coupling in perturbative quantum field theory is scheme-dependent.  As discussed by Grunberg~\cite{Grunberg},  an effective coupling or charge can be defined directly from physical observables.
Effective charges defined from
different observables can be related  to each other in the leading-twist domain using commensurate scale relations
 (CSR)~\cite{CSR}.    The  potential between infinitely heavy quarks can be defined analytically in momentum transfer
 space as the product  of the running coupling times the Born gluon propagator: $V(q)  = - 4 \pi C_F {\alpha_V(q) / q^2}$.   This effective charge defines a renormalization scheme -- the $\alpha_V$ scheme of Appelquist, Dine,  and Muzinich~\cite{Appelquist:1977tw}.
In fact, the holographic coupling $\alpha_s^{AdS}(Q^2)$ can be considered to be the nonperturbative extension of the
$\alpha_V$ effective charge defined in Ref. \cite{Appelquist:1977tw}.
We can also make  use of the $g_1$ scheme, where the strong coupling $\alpha_{g_1}(Q^2)$ is determined from
the Bjorken sum rule~\cite{BjorkenSR}. The coupling $\alpha_{g_1}(Q^2)$ has the advantage that it is the best-measured effective charge, and it can be used to extrapolate the definition of the effective coupling to large distances~\cite{Deur:2009hu}.  Since $\alpha_{g_1}$ has been measured at intermediate energies, it is
particularly useful for studying  the transition from
small to large distances.

We have recently  shown~\cite{Brodsky:2010ur}   how the LF holographic mapping of effective classical gravity in AdS space, modified by a positive-sign dilaton background, can  be used to identify an analytically simple  color-confining
non-perturbative effective coupling $\alpha_s^{AdS}(Q^2)$ as a function of the space-like momentum transfer $Q^2 = - q^2$.   This coupling incorporates  confinement
and agrees well with effective charge observables and lattice simulations.
It also exhibits an infrared fixed point at small $Q^2$ and asymptotic freedom at large $Q^2$. However, the fall-off   of
$\alpha_s^{AdS}(Q^2)$  at large $Q^2$ is exponential: $\alpha_s^{AdS}(Q^2) \sim e^{-Q^2 /  \kappa^2}$, rather than the
pQCD logarithmic fall-off.    It agrees with hadron physics data
extracted phenomenologically from different observables, as well as with  the predictions of models with built-in confinement  and lattice simulations. We
also show that a phenomenological extended coupling can be defined which implements the pQCD behavior.
The  $\beta$-function derived from light-front holography becomes significantly  negative in the non-perturbative regime $Q^2 \sim \kappa^2$, where it reaches a minimum, signaling the transition region from the infrared (IR) conformal region, characterized by hadronic degrees of freedom,  to a pQCD conformal ultraviolet (UV)  regime where the relevant degrees of freedom are the quark and gluon constituents.  The  $\beta$-function vanishes at large $Q^2$ consistent with asymptotic freedom, and it vanishes at small $Q^2$ consistent with an infrared fixed point~\cite{Brodsky:2008be, Cornwall:1981zr}.

Let us consider a five-dimensional gauge field $F$ propagating in AdS$_5$ space in presence of a dilaton background
$\varphi(z)$ which introduces the energy scale $\kappa$ in the five-dimensional action.
At quadratic order in the field strength the action is
\begin{equation}
S =  - {1\over 4}\int \! d^5x \, \sqrt{g} \, e^{\varphi(z)}  {1\over g^2_5} \, F^2,
\label{eq:action}
\end{equation}
where the metric determinant of AdS$_5$ is $\sqrt g = ( {R/z})^5$,  $\varphi=  \kappa^2 z^2$ and the square of the coupling $g_5$ has dimensions of length.   We  can identify the prefactor
\begin{equation} \label{eq:flow}
g^{-2}_5(z) =  e^{\varphi(z)}  g^{-2}_5 ,
\end{equation}
in the  AdS  action (\ref{eq:action})  as the effective coupling of the theory at the length scale $z$.
The coupling $g_5(z)$ then incorporates the non-conformal dynamics of confinement. The five-dimensional coupling $g_5(z)$
is mapped,  modulo a  constant, into the Yang-Mills (YM) coupling $g_{YM}$ of the confining theory in physical space-time using light-front holography. One  identifies $z$ with the invariant impact separation variable $\zeta$ which appears in the LF Hamiltonian:
$g_5(z) \to g_{YM}(\zeta)$. Thus
\begin{equation}  \label{eq:gYM}
\alpha_s^{AdS}(\zeta) = g_{YM}^2(\zeta)/4 \pi \propto  e^{-\kappa^2 \zeta^2} .
\end{equation}

In contrast with the 3-dimensional radial coordinates of the non-relativistic Schr\"o-dinger theory, the natural light-front
variables are the two-dimensional cylindrical coordinates $(\zeta, \phi)$ and the
light-cone fraction $x$. The physical coupling measured at the scale $Q$ is the two-dimensional Fourier transform
of the  LF transverse coupling $\alpha_s^{AdS}(\zeta)$  (\ref{eq:gYM}). Integration over the azimuthal angle
 $\phi$ gives the Bessel transform
 \begin{equation} \label{eq:2dimFT}
\alpha_s^{AdS}(Q^2) \sim \int^\infty_0 \! \zeta d\zeta \,  J_0(\zeta Q) \, \alpha_s^{AdS}(\zeta),
\end{equation}
in the $q^+ = 0$ light-front frame where $Q^2 = -q^2 = - \mbf{q}_\perp^2 > 0$ is the square of the space-like
four-momentum transferred to the
hadronic bound state.   Using this ansatz we then have from  Eq.  (\ref{eq:2dimFT})
\begin{equation}
\label{eq:alphaAdS}
\alpha_s^{AdS}(Q^2) = \alpha_s^{AdS}(0) \, e^{- Q^2 /4 \kappa^2}.
\end{equation}
In contrast, the negative dilaton solution $\varphi=  -\kappa^2 z^2$ leads to an integral which diverges at large $\zeta$.
We identify $\alpha_s^{AdS}(Q^2)$ with the physical QCD running coupling
in its nonperturbative domain.

The flow equation  (\ref{eq:flow}) from the scale dependent measure for the gauge fields can be understood as a consequence of field-strength renormalization.
In physical QCD we can rescale the non-Abelian gluon field  $A^\mu \to \lambda A^\mu$  and field strength
$G^{\mu \nu}  \to \lambda G^{\mu \nu}$  in the QCD Lagrangian density  $\mathcal{L}_{\rm QCD}$ by a compensating rescaling of the coupling strength $g \to \lambda^{-1} g.$  The renormalization of the coupling $g _{phys} = Z^{1/2}_3  g_0,$   where $g_0$ is the bare coupling in the Lagrangian in the UV-regulated theory,  is thus  equivalent to the renormalization of the vector potential and field strength: $A^\mu_{ren} =  Z_3^{-1/2} A^\mu_0$, $G^{\mu \nu}_{ren} =  Z_3^{-1/2} G^{\mu \nu}_0$   with a rescaled Lagrangian density
${\cal L}_{\rm QCD}^{ren}  = Z_3^{-1}  { \cal L}_{\rm QCD}^0  = (g_{phys}/g_0)^{-2}  \mathcal{L}_0$.
 In lattice gauge theory,  the lattice spacing $a$ serves as the UV regulator, and the renormalized QCD coupling is determined  from the normalization of the gluon field strength as it appears  in the gluon propagator. The inverse of the lattice size $L$ sets the mass scale of the resulting running coupling.
As is the case in lattice gauge theory, color confinement in AdS/QCD reflects nonperturbative dynamics at large distances. The QCD couplings defined from lattice gauge theory and the soft wall holographic model are thus similar in concept, and both schemes are expected to have similar properties in the nonperturbative domain, up to a rescaling of their respective momentum scales.

\subsection{Comparison of  the Holographic Coupling with Other Effective Charges \label{alphatest}}

The effective coupling  $\alpha^{AdS}(Q^2)$ (solid line) is compared in Fig. \ref{alphas} with  experimental and lattice data. For this comparison to be meaningful, we have to impose the same normalization on the AdS coupling as the $g_1$ coupling. This defines $\alpha_s^{AdS}$ normalized to the $g_1$ scheme: $\alpha_{g_1}^{AdS}\left(Q^2 \! =0\right) = \pi.$
Details on the comparison with other effective charges are given in Ref. ~\cite{Deur:2005cf}.

\begin{figure}[h]
\begin{center}
\includegraphics[width=7.0cm]{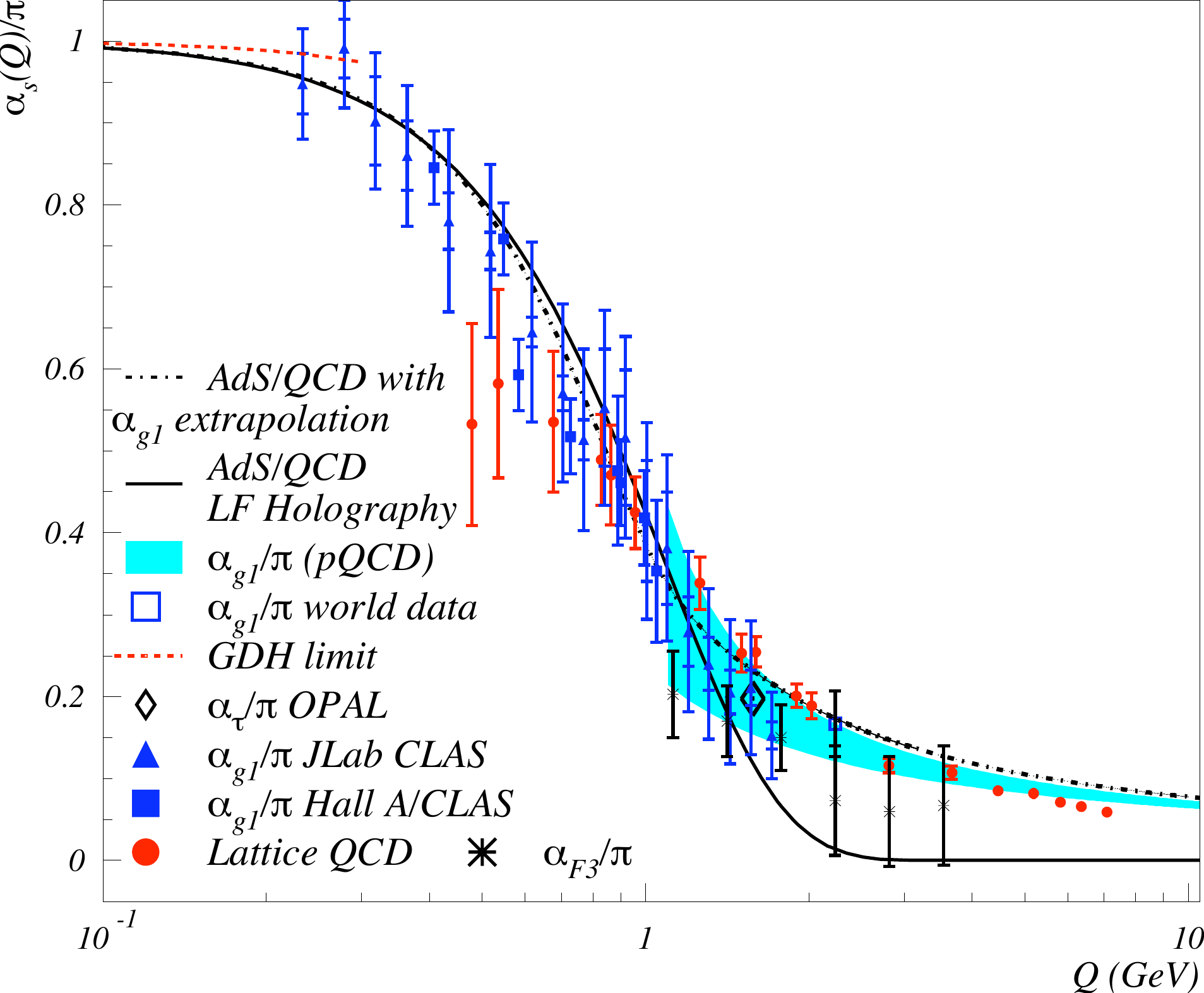} ~~~
\includegraphics[width=6.8cm]{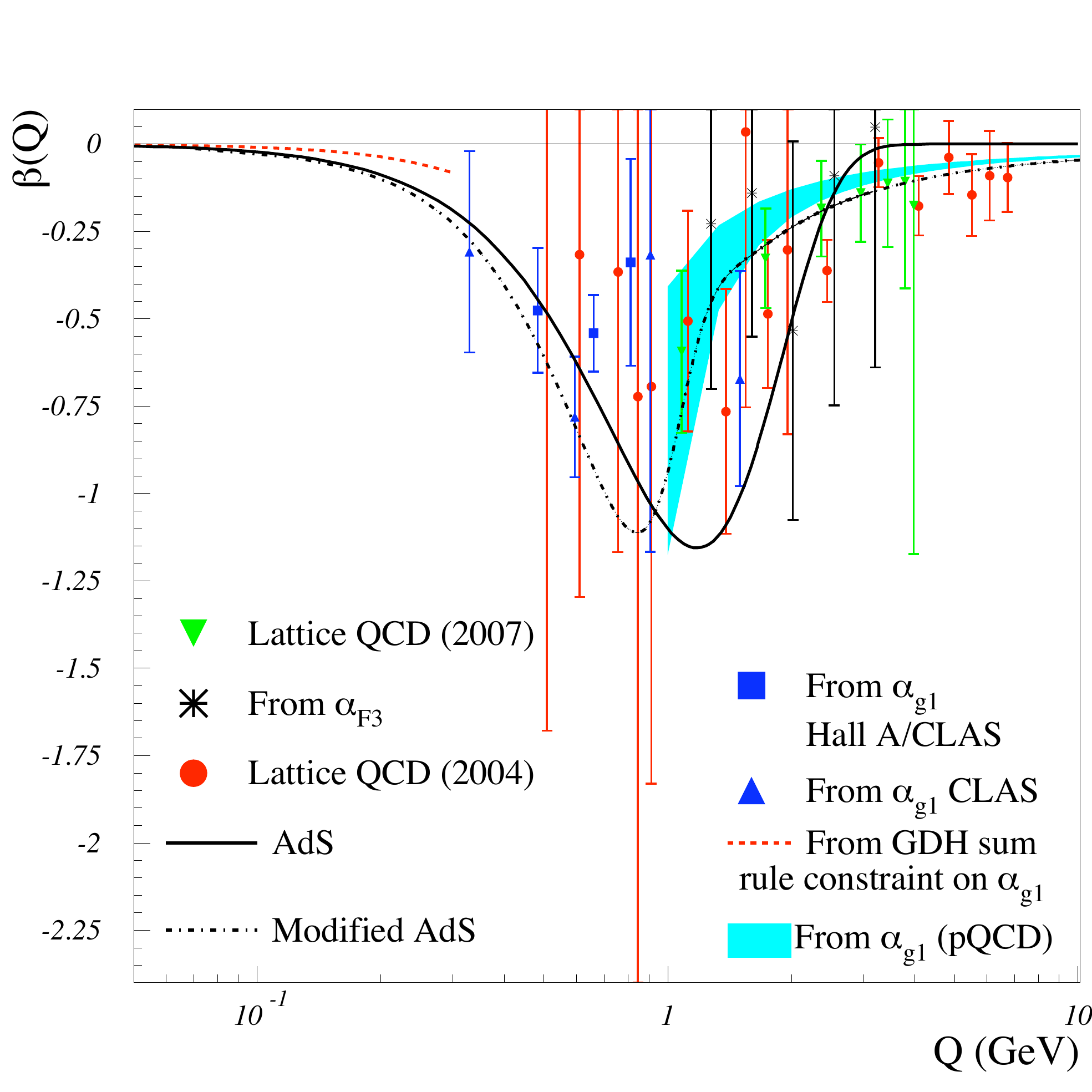}
 \caption{(a) Effective coupling  from LF holography  for  $\kappa = 0.54 ~ {\rm GeV}$ compared with effective QCD couplings  extracted from
different observables and lattice results. (b) Prediction for the $\beta$-function compared to  lattice simulations,  JLab and CCFR results  for the Bjorken sum rule effective charge.}
\label{alphas}
\end{center}
\end{figure}

The couplings in Fig. \ref{alphas} (a) agree well in the strong coupling regime  up to $Q  \! \sim \! 1$ GeV.  The value $\kappa  = 0.54 ~ {\rm GeV}$ is determined from the vector meson Regge trajectory~\cite {deTeramond:2009xk}.
The lattice results shown in Fig. \ref{alphas} from Ref.~\cite{Furui:2006py} have been scaled to match the perturbative UV domain. The effective charge $\alpha_{ g_1}$ has been determined in Ref.~\cite{Deur:2005cf} from several experiments.  Fig. \ref{alphas} also displays other couplings from different observables as well as $\alpha_{g_1}$ which is computed from the
Bjorken sum rule~\cite{BjorkenSR}  over a large range of momentum transfer (cyan band). At $Q^2\!=\!0$ one has the constraint  on the slope of $\alpha_{g_1}$ from the Gerasimov-Drell-Hearn (GDH) sum rule~\cite{GDH} which is also shown in the figure.
The results show no sign of a phase transition, cusp, or other non-analytical behavior, a fact which allows us to extend the functional dependence of the coupling to large distances.
As discussed below,
the smooth behavior of the  AdS strong
coupling also allows us to extrapolate its form to the perturbative domain~\cite{Brodsky:2010ur}.

The hadronic model obtained from the dilaton-modified AdS space provides a semi-classical first approximation to QCD.  Color confinement is introduced by the harmonic oscillator potential, but effects from gluon creation and absorption are not included in this effective theory.  The nonperturbative  confining effects vanish exponentially at large momentum transfer (Eq. (\ref{eq:alphaAdS})), and thus the logarithmic fall-off from pQCD quantum loops will dominate in this regime.
The  running coupling  $\alpha_s^{AdS}$ given by Eq.  (\ref{eq:alphaAdS})  is obtained from a
color-confining potential.   Since the strong coupling is an analytical function of the momentum transfer at all scales, we can extend  the range of applicability of $\alpha_s^{AdS}$ by matching to a  perturbative coupling
at the transition scale  $Q \sim 1$ GeV, where pQCD contributions become important, as described in Ref. ~\cite{Brodsky:2010ur}.
The smoothly extrapolated result (dot-dashed line) for  $\alpha_{s}$ is also shown on Fig.~\ref{alphas}.
In order to have a fully analytical model, we write
\begin{equation}
\label{eq:alphafit}
\alpha_{Modified, g_1}^{AdS}(Q^2) = \alpha_{g_1}^{AdS}(Q^2) g_+(Q^2 ) + \alpha_{g_1}^{fit}(Q^2) g_-(Q^2),
\end{equation}
where $g_{\pm}(Q^2) = 1/(1+e^{\pm \left(Q^2 - Q^2_0\right)/\tau^2})$ are smeared step functions which match the two
regimes. The parameter $\tau$ represents the width of the
transition region.    Here $\alpha_{g_1}^{AdS}$ is given by Eq. (\ref{eq:alphaAdS}) with the normalization  $\alpha_{g_1}^{AdS}(0)=\pi$
-- the plain black line in Fig.~\ref{alphas} -- and $\alpha_{g_{1}}^{fit}$ in Eq. (\ref{eq:alphafit}) is the analytical
fit to the measured coupling $\alpha_{g_1}$~\cite{Deur:2005cf}.
The couplings  are chosen to have the same  normalization at $Q^2=0.$
The smoothly extrapolated result (dot-dashed line) for  $\alpha_{s}$ is also shown on Fig.~\ref{alphas}. We use the
parameters $Q_{0}^{2}=0.8$ GeV$^{2}$ and $\tau^2=0.3$ GeV$^{2}$.

\subsection{The $\beta$-Function from AdS/QCD}

The $\beta$-function for the nonperturbative  effective coupling obtained from the LF holographic mapping in a positive dilaton modified AdS background  is
\begin{equation} \label{eq:beta}
\beta^{AdS}(Q^2)  = {d \over d \log{Q^2}}\alpha^{AdS}(Q^2) = - {\pi Q^2\over 4 \kappa^2} e^{-Q^2/(4 \kappa^2)}.
\end{equation}
The solid line in Fig. \ref{alphas} (b) corresponds to the light-front holographic result Eq.  (\ref{eq:beta}).    Near $Q_0 \simeq 2 \kappa \simeq 1$ GeV, we can interpret the results as a transition from  the nonperturbative IR domain to the quark and gluon degrees of freedom in the perturbative UV  regime. The transition momentum  scale $Q_0$  is compatible with the momentum transfer for the onset of scaling behavior in exclusive reactions where quark counting rules are observed~\cite{Brodsky:1973kr}.
For example, in deuteron photo-disintegration the onset of scaling corresponds to  momentum transfer  of  1.0  GeV to the nucleon involved~\cite{Gao:2004zh}.  Dimensional counting is built into the AdS/QCD soft and hard wall models since the AdS amplitudes $\Phi(z)$ are governed by their twist scaling behavior $z^\tau$ at short distances, $ z \to 0$~\cite{Polchinski:2001tt}.

Also shown on Fig. \ref{alphas} (b) are the $\beta$-functions obtained from phenomenology and lattice calculations. For clarity, we present only the LF holographic predictions, the lattice results from  \cite{Furui:2006py}, and the
experimental data supplemented by the relevant sum rules.
The dot-dashed curve corresponds to the
extrapolated approximation obtained by matching to AdS results to the perturbative coupling~\cite{Brodsky:2010ur}
given by Eq. (\ref{eq:alphafit}).
The $\beta$-function extracted from LF holography, as well as the forms obtained from
the works of Cornwall~\cite{Cornwall:1981zr}, Bloch, Fisher {\it et al.}~\cite{S-Deq.}, Burkert and Ioffe~\cite{Burkert-Ioffe} and Furui and Nakajima~\cite{Furui:2006py}, are seen to have a similar shape
and magnitude.

Judging from these results, we infer that the   actual  $\beta$-function of QCD will extrapolate between the non-perturbative results for $Q < 1$ GeV and the pQCD results
for $Q > 1$ GeV. We also observe that the general conditions
\begin{eqnarray}
& \beta(Q \to 0) =  \beta(Q \to \infty) = 0 , \label{a} \\
&  \beta(Q)  <  0, ~ {\rm for} ~  Q > 0 , \label{b}\\
& \frac{d \beta}{d Q} \big \vert_{Q = Q_0}  = 0, \label{c} \\
& \frac{d \beta}{d Q}   < 0, ~ {\rm for} ~ Q < Q_0, ~~
 \frac{d \beta}{d Q}   > 0, ~ {\rm for} ~ Q > Q_0 \label{d} .
\end{eqnarray}
are satisfied by our model $\beta$-function obtained from LF holography.

Eq. (\ref{a}) expresses the fact that  QCD approaches a conformal theory in both the far ultraviolet and deep infrared regions. In the semiclassical approximation to QCD  without particle creation or absorption,
the $\beta$-function is zero and the approximate theory is scale  invariant
in the limit of massless quarks~\cite{Parisi:1972zy}. When quantum corrections are included,
the conformal behavior is
preserved at very large $Q$ because of asymptotic freedom and near $Q \to 0$ because the theory develops a fixed
point.  An infrared fixed point is in fact a natural consequence of color confinement~\cite{Cornwall:1981zr}:
since the propagators of the colored fields have a maximum wavelength,  all loop
integrals in the computation of  the gluon self-energy decouple at $Q^2 \to 0$~\cite{Brodsky:2008be}. Condition (\ref{b}) for $Q^2$ large, expresses the basic anti-screening behavior of QCD where the strong coupling vanishes. The $\beta$-function in QCD is essentially negative, thus the coupling increases monotonically from the UV to the IR where it reaches its maximum value:  it has a finite value for a theory with a mass gap. Equation (\ref{c}) defines the transition region at $Q_0$ where the 
$\beta$-function has a minimum.  Since there is only one hadronic-partonic transition, the minimum is an absolute minimum; thus the additional conditions expressed in Eq (\ref{d}) follow immediately from
Eqs.  (\ref{a}-\ref{c}). The conditions given by Eqs.  (\ref{a}-\ref{d}) describe the essential
behavior of the full $\beta$-function for an effective QCD coupling whose scheme/definition is similar to that of the $V$-scheme.

\section{Conclusions \label{conclusions}}

The combination of Anti-de Sitter  space
methods with light-front  holography provides an accurate first approximation for the spectra and wavefunctions of meson and baryon light-quark  bound states. One also obtains an elegant connection between a semiclassical first approximation to QCD, quantized on the light-front, with hadronic modes propagating on a fixed AdS background. The resulting bound-state Hamiltonian equation of motion in QCD leads to  relativistic light-front wave equations in the invariant impact variable $\zeta$ which measures the separation of the quark and gluonic constituents within the hadron at equal light-front time. This corresponds
to the effective single-variable relativistic Schr\"odinger-like equation in the AdS fifth dimension coordinate $z$,  Eq. (\ref{eq:QCDLFWE}). The eigenvalues give the hadronic spectrum, and the eigenmodes represent the probability distributions of the hadronic constituents at a given scale.
In particular, the light-front holographic mapping of  effective classical gravity in AdS space, modified by a positive-sign dilaton background,  provides a very good description of the spectrum and form factors of light mesons and baryons.
We have also shown that the light-front holographic mapping of  effective classical gravity in AdS space, modified by the positive-sign dilaton background  predicts the form of a non-perturbative effective coupling $\alpha_s^{AdS}(Q)$ and its $\beta$-function.
The AdS/QCD running coupling is in very good agreement with the effective
coupling $\alpha_{g_1}$ extracted from  the Bjorken sum rule.
The holographic $\beta$-function displays a transition from  nonperturbative to perturbative  regimes  at a momentum scale $Q \sim 1$ GeV.
Our analysis indicates that light-front holography captures the characteristics of the full  $\beta$-function of QCD  and  the essential dynamics of confinement, thus giving further support to the application of the gauge/gravity duality to the confining dynamics of strongly coupled QCD.

There are many phenomenological applications where detailed knowledge of the QCD coupling and the renormalized gluon propagator at relatively soft momentum transfer are essential.
This includes exclusive and semi-exclusive processes as well as the rescattering interactions which
create the leading-twist Sivers single-spin correlations in
semi-inclusive deep inelastic scattering~\cite{Brodsky:2002cx, Collins:2002kn}, the Boer-Mulders functions which lead to anomalous  $\cos 2 \phi$  contributions to the lepton pair angular distribution in the unpolarized Drell-Yan reaction~\cite{Boer:2002ju}, and the Sommerfeld-Sakharov-Schwinger correction to heavy quark production at
threshold~\cite{Brodsky:1995ds}.
The confining AdS/QCD coupling from light-front holography thus can provide a
quantitative understanding of this factorization-breaking physics~\cite{Collins:2007nk}.

\newpage

\noindent{\bf Acknowledgments}

\vspace{5pt}

We thank  Volker Burkert, John Cornwall, Sadataka Furui, 
Philipp H\"agler, Wolfgang Korsch, G. Peter Lepage, Takemichi Okui, Joannis Papavassiliou, Anatoly Radyushkin, Craig Roberts, Robert Shrock, and Peter Tandy for helpful comments.
This research was supported by the Department of Energy contracts DE--AC02--76SF00515 and  DE-AC05-84ER40150.
Invited talk, presented by SJB at the
Workshop on Exclusive Reactions at High Momentum Transfer (IV),
May 18-21, 2010,
Thomas Jefferson National Accelerator Facility,
Newport News, VA.


\begin{thebibliography}{9}

\bibitem{Brodsky:1973kr}
  S.~J.~Brodsky and G.~R.~Farrar,
  Phys.\ Rev.\ Lett.\  {\bf 31}, 1153 (1973);
  Phys.\ Rev.\  D {\bf 11}, 1309 (1975).

\bibitem{Matveev:1973ra}
  V.~A.~Matveev, R.~M.~Muradian and A.~N.~Tavkhelidze,
  Lett.\ Nuovo Cim.\  {\bf 7}, 719 (1973).

  \bibitem{Lepage:1979zb}
  G.~P.~Lepage and S.~J.~Brodsky,
  Phys.\ Lett.\  B {\bf 87}, 359 (1979).

\bibitem{Efremov:1979qk}
  A.~V.~Efremov and A.~V.~Radyushkin,
  Phys.\ Lett.\  B {\bf 94}, 245 (1980).

\bibitem{Polchinski:2001tt}
  J.~Polchinski and M.~J.~Strassler,
  Phys.\ Rev.\ Lett.\  {\bf 88}, 031601 (2002)
  [arXiv:hep-th/0109174].
  
  \bibitem{Brodsky:1988xz}
  S.~J.~Brodsky and A.~H.~Mueller,
  Phys.\ Lett.\  B {\bf 206}, 685 (1988).

\bibitem{Bertsch:1981py}
  G.~Bertsch, S.~J.~Brodsky, A.~S.~Goldhaber and J.~F.~Gunion,
  Phys.\ Rev.\ Lett.\  {\bf 47}, 297 (1981).

\bibitem{Gunion:1973ex}
  J.~F.~Gunion, S.~J.~Brodsky, R.~Blankenbecler,
  Phys.\ Rev.\  {\bf D8}, 287 (1973).

\bibitem{Sivers:1975dg}
  D.~W.~Sivers, S.~J.~Brodsky and R.~Blankenbecler,
  Phys.\ Rept.\  {\bf 23}, 1 (1976).

\bibitem{White:1994tj}
  C.~G.~White {\it et al.},
  Phys.\ Rev.\  D {\bf 49}, 58 (1994).

\bibitem{Holt:1990ze}
  R.~J.~Holt,
  Phys.\ Rev.\  C {\bf 41}, 2400 (1990).

\bibitem{Bochna:1998ca}
  C.~Bochna {\it et al.}  [E89-012 Collaboration],
  Phys.\ Rev.\ Lett.\  {\bf 81}, 4576 (1998)
  [arXiv:nucl-ex/9808001].

\bibitem{Rossi:2004qm}
  P.~Rossi {\it et al.}  [CLAS Collaboration],
  Phys.\ Rev.\ Lett.\  {\bf 94}, 012301 (2005)
  [arXiv:hep-ph/0405207].

\bibitem{Rock:1991jy}
  S.~Rock {\it et al.},
  Phys.\ Rev.\  D {\bf 46}, 24 (1992).

\bibitem{Brodsky:1976rz}
  S.~J.~Brodsky and B.~T.~Chertok,
  Phys.\ Rev.\  D {\bf 14}, 3003 (1976).

\bibitem{Brodsky:1983vf}
  S.~J.~Brodsky, C.~R.~Ji and G.~P.~Lepage,
  Phys.\ Rev.\ Lett.\  {\bf 51}, 83 (1983).

\bibitem{Brodsky:1989pv}
  S.~J.~Brodsky and G.~P.~Lepage,
  Adv.\ Ser.\ Direct.\ High Energy Phys.\  {\bf 5}, 93 (1989).

\bibitem{Aitala:2000hc}
  E.~M.~Aitala {\it et al.}  [E791 Collaboration],
  Phys.\ Rev.\ Lett.\  {\bf 86}, 4773 (2001)
  [arXiv:hep-ex/0010044].

\bibitem{Clasie:2007gq}
  B.~Clasie {\it et al.},
  arXiv:0707.1481 [nucl-ex].

\bibitem{Airapetian:2002eh}
  A.~Airapetian {\it et al.}  [HERMES Collaboration],
  Phys.\ Rev.\ Lett.\  {\bf 90}, 052501 (2003)
  [arXiv:hep-ex/0209072].

\bibitem{Berger:1979du}
  E.~L.~Berger, S.~J.~Brodsky,
  Phys.\ Rev.\ Lett.\  {\bf 42}, 940-944 (1979).

\bibitem{Berger:1980qg}
  E.~L.~Berger, T.~Gottschalk, D.~W.~Sivers,
  Phys.\ Rev.\  {\bf D23}, 99 (1981).

\bibitem{Berger:1981fr}
  E.~L.~Berger, S.~J.~Brodsky,
  Phys.\ Rev.\  {\bf D24}, 2428 (1981).

\bibitem{Arleo:2010yg}
  F.~Arleo, S.~J.~Brodsky, D.~S.~Hwang {\it et al.},
  [arXiv:1006.4045 [hep-ph]].

\bibitem{Arleo:2009ch}
  F.~Arleo, S.~J.~Brodsky, D.~S.~Hwang {\it et al.},
  [arXiv:0911.4604 [hep-ph]].

\bibitem{Brodsky:2008qp}
  S.~J.~Brodsky, A.~Sickles,
  Phys.\ Lett.\  {\bf B668}, 111-115 (2008)
  [arXiv:0804.4608 [hep-ph]].

\bibitem{Druzhinin:2009gq}
  V.~P.~Druzhinin,
  PoS {\bf EPS-HEP2009}, 051 (2009)
  [arXiv:0909.3148 [hep-ex]] and preliminary results presented at IHCEP2010.

\bibitem{Radyushkin:2009zg}
  A.~V.~Radyushkin,
  Phys.\ Rev.\  {\bf D80}, 094009 (2009)
  [arXiv:0906.0323 [hep-ph]].

\bibitem{Polyakov:2009je}
  M.~V.~Polyakov,
  JETP Lett.\  {\bf 90}, 228-231 (2009)
  [arXiv:0906.0538 [hep-ph]].

\bibitem{Mikhailov:2009sa}
  S.~V.~Mikhailov, N.~G.~Stefanis,
  Mod.\ Phys.\ Lett.\  {\bf A24}, 2858-2867 (2009)
  [arXiv:0910.3498 [hep-ph]].

\bibitem{Brodsky:1981rp}
  S.~J.~Brodsky, G.~P.~Lepage,
  Phys.\ Rev.\  {\bf D24}, 1808 (1981).

\bibitem{Danagoulian:2007gs}
  A.~Danagoulian {\it et al.}  [Hall A Collaboration],
  Phys.\ Rev.\ Lett.\  {\bf 98}, 152001 (2007)
  [arXiv:nucl-ex/0701068].

\bibitem{Diehl:1998kh}
  M.~Diehl, T.~Feldmann, R.~Jakob and P.~Kroll,
  Eur.\ Phys.\ J.\  C {\bf 8}, 409 (1999)
  [arXiv:hep-ph/9811253].

\bibitem{Chen:2001sm}
  A.~Chen,
{\it
International Conference on the Structure and Interactions of the Photon and 14th International Workshop on Photon-Photon
Collisions (Photon 2001), Ascona, Switzerland, 2-7 Sep 2001}.

\bibitem{Court:1986dh}
  G.~R.~Court {\it et al.},
  Phys.\ Rev.\ Lett.\  {\bf 57}, 507 (1986).

\bibitem{Mardor:1998zf}
  I.~Mardor {\it et al.},
  Phys.\ Rev.\ Lett.\  {\bf 81}, 5085 (1998).

\bibitem{Brodsky:1987xw}
  S.~J.~Brodsky and G.~F.~de Teramond,
  Phys.\ Rev.\ Lett.\  {\bf 60}, 1924 (1988).
  
  \bibitem{Dutta:2004fw}
  D.~Dutta and H.~Gao,
  Phys.\ Rev.\  C {\bf 71}, 032201 (2005)
  [arXiv:hep-ph/0411267].
  
  \bibitem{Brodsky:1997fj}
  S.~J.~Brodsky and M.~Karliner,
  Phys.\ Rev.\ Lett.\  {\bf 78}, 4682 (1997)
  [arXiv:hep-ph/9704379].

\bibitem{Dirac:1949cp}
  P.~A.~M.~Dirac,
  Rev.\ Mod.\ Phys.\  {\bf 21}, 392 (1949).
  
\bibitem{Pauli:1985ps}
  H.~C.~Pauli and S.~J.~Brodsky,
  Phys.\ Rev.\  D {\bf 32}, 2001 (1985).

\bibitem{Brodsky:1997de}
  S.~J.~Brodsky, H.~C.~Pauli and S.~S.~Pinsky,
  Phys.\ Rept.\  {\bf 301}, 299 (1998)
  [arXiv:hep-ph/9705477].

  \bibitem{Drell:1969km}
  S.~D.~Drell and T.~M.~Yan,
  Phys.\ Rev.\ Lett.\  {\bf 24}, 181 (1970).

\bibitem{West:1970av}
  G.~B.~West,
  Phys.\ Rev.\ Lett.\  {\bf 24}, 1206 (1970).

\bibitem{Brodsky:1998hn}
  S.~J.~Brodsky and D.~S.~Hwang,
  Nucl.\ Phys.\  B {\bf 543}, 239 (1999)
  [arXiv:hep-ph/9806358].

\bibitem{Brodsky:2000xy}
  S.~J.~Brodsky, M.~Diehl and D.~S.~Hwang,
  Nucl.\ Phys.\  B {\bf 596}, 99 (2001)
  [arXiv:hep-ph/0009254].

\bibitem{Lepage:1980fj}
  G.~P.~Lepage and S.~J.~Brodsky,
  Phys.\ Rev.\  D {\bf 22}, 2157 (1980).
  
  \bibitem{Brodsky:1979qm}
  S.~J.~Brodsky and G.~P.~Lepage,
  SLAC-PUB-2294;
Workshop on Current Topics in High Energy Physics, Cal Tech., Pasadena, Calif., Feb 13-17, 1979.

  \bibitem{Teryaev:1999su}
  O.~V.~Teryaev,
  arXiv:hep-ph/9904376.

   \bibitem{Brodsky:2000ii}
  S.~J.~Brodsky, D.~S.~Hwang, B.~Q.~Ma and I.~Schmidt,
  Nucl.\ Phys.\  B {\bf 593}, 311 (2001)
  [arXiv:hep-th/0003082].
  
   \bibitem{Brodsky:2009zd}
  S.~J.~Brodsky and R.~Shrock,
  arXiv:0905.1151 [hep-th].
  S.~J.~Brodsky, C.~D.~Roberts, R.~Shrock, and P.~Tandy
   [arXiv:1005.4610 [nucl-th]].

\bibitem{Maldacena:1997re}
  J.~M.~Maldacena,
  Adv.\ Theor.\ Math.\ Phys.\  {\bf 2}, 231 (1998)
  [Int.\ J.\ Theor.\ Phys.\  {\bf 38}, 1113 (1999)]
  [arXiv:hep-th/9711200].

\bibitem{Furui:2006py}
  S.~Furui and H.~Nakajima,
  Phys.\ Rev.\  D {\bf 76}, 054509 (2007)
  [arXiv:hep-lat/0612009].

\bibitem{vonSmekal:1997is}
  L.~von Smekal, R.~Alkofer and A.~Hauck,
  Phys.\ Rev.\ Lett.\  {\bf 79}, 3591 (1997)
  [arXiv:hep-ph/9705242].

  \bibitem{Deur:2005cf}
  A.~Deur, V.~Burkert, J.~P.~Chen and W.~Korsch,
  Phys.\ Lett.\  B {\bf 650}, 244 (2007)
  [arXiv:hep-ph/0509113].

\bibitem{Deur:2008rf}
  A.~Deur, V.~Burkert, J.~P.~Chen and W.~Korsch,
  Phys.\ Lett.\  B {\bf 665}, 349 (2008)
  [arXiv:0803.4119 [hep-ph]].

\bibitem{Brodsky:2008be}
  S.~J.~Brodsky and R.~Shrock,
  Phys.\ Lett.\  B {\bf 666}, 95 (2008)
  [arXiv:0806.1535 [hep-th]].

\bibitem{Karch:2006pv}
  A.~Karch, E.~Katz, D.~T.~Son and M.~A.~Stephanov,
  Phys.\ Rev.\  D {\bf 74}, 015005 (2006)
  [arXiv:hep-ph/0602229].

\bibitem{deTeramond:2009xk}
  G.~F.~de Teramond and S.~J.~Brodsky,
  Nucl.\ Phys.\ Proc.\ Suppl.\  {\bf 199}, 89 (2010)
  [arXiv:0909.3900 [hep-ph]].

\bibitem{Andreev:2006ct}
  O.~Andreev and V.~I.~Zakharov,
  Phys.\ Rev.\  D {\bf 74}, 025023 (2006)
  [arXiv:hep-ph/0604204].

\bibitem{Zuo:2009dz}
  F.~Zuo,
  arXiv:0909.4240 [hep-ph].

\bibitem{Afonin:2010fr}
  S.~S.~Afonin,
  arXiv:1001.3105 [hep-ph].

\bibitem{Glazek:1987ic}
  S.~D.~Glazek and M.~Schaden,
  Phys.\ Lett.\  B {\bf 198}, 42 (1987).

\bibitem{Hoyer:2009ep}
  P.~Hoyer,
  arXiv:0909.3045 [hep-ph].

\bibitem{deTeramond:2008ht}
  G.~F.~de Teramond and S.~J.~Brodsky,
  Phys.\ Rev.\ Lett.\  {\bf 102}, 081601 (2009)
  [arXiv:0809.4899 [hep-ph]].

\bibitem{Brodsky:2006uqa}
  S.~J.~Brodsky and G.~F.~de Teramond,
  Phys.\ Rev.\ Lett.\  {\bf 96}, 201601 (2006)
  [arXiv:hep-ph/0602252].

\bibitem{Brodsky:2007hb}
  S.~J.~Brodsky and G.~F.~de Teramond,
  Phys.\ Rev.\  D {\bf 77}, 056007 (2008)
  [arXiv:0707.3859 [hep-ph]].

\bibitem{Brodsky:2008pf}
  S.~J.~Brodsky and G.~F.~de Teramond,
  Phys.\ Rev.\  D {\bf 78}, 025032 (2008)
  [arXiv:0804.0452 [hep-ph]].

\bibitem{deTeramond:2010we}
  G.~F.~de Teramond and S.~J.~Brodsky,
  arXiv:1001.5193 [hep-ph].

\bibitem{Abidin:2008ku}
  Z.~Abidin and C.~E.~Carlson,
  Phys.\ Rev.\  D {\bf 77}, 095007 (2008)
  [arXiv:0801.3839 [hep-ph]].

\bibitem{Amsler:2008xx}
  C. Amsler {\it et al.}  (Particle Data Group),
  Phys.\ Lett.\  B {\bf 667}, 1, (2008).

    \bibitem{Baldini:1998qn}
  R.~Baldini, S.~Dubnicka, P.~Gauzzi, S.~Pacetti, E.~Pasqualucci and Y.~Srivastava,
  Eur.\ Phys.\ J.\  C {\bf 11}, 709 (1999).

  \bibitem{Tadevosyan:2007yd}
  V.~Tadevosyan {\it et al.}  [Jefferson Lab F(pi) Collaboration],
  Phys.\ Rev.\  C {\bf 75}, 055205 (2007)
  [arXiv:nucl-ex/0607007].

  \bibitem{Horn:2006tm}
  T.~Horn {\it et al.}  [Fpi2 Collaboration],
  Phys.\ Rev.\ Lett.\  {\bf 97}, 192001 (2006)
  [arXiv:nucl-ex/0607005].

\bibitem{Craig:2009rk}
  N.~J.~Craig and D.~Green,
  JHEP {\bf 0909}, 113 (2009)
  [arXiv:0905.4088 [hep-ph]].

\bibitem{Brodsky:2003px}
  S.~J.~Brodsky and G.~F.~de Teramond,
  Phys.\ Lett.\  B {\bf 582}, 211 (2004)
  [arXiv:hep-th/0310227].

\bibitem{Erlich:2005qh}
  J.~Erlich, E.~Katz, D.~T.~Son and M.~A.~Stephanov,
  Phys.\ Rev.\ Lett.\  {\bf 95}, 261602 (2005)
  [arXiv:hep-ph/0501128].

   \bibitem{DaRold:2005zs}
  L.~Da Rold and A.~Pomarol,
  Nucl.\ Phys.\ B {\bf 721}, 79 (2005)
  [arXiv:hep-ph/0501218];
  JHEP {\bf 0601}, 157 (2006)
  [arXiv:hep-ph/0510268].

\bibitem{Klempt:2007cp}
  E.~Klempt and A.~Zaitsev,
  Phys.\ Rept.\  {\bf 454}, 1 (2007)
  [arXiv:0708.4016 [hep-ph]].

\bibitem{BoschiFilho:2005yh}
  H.~Boschi-Filho, N.~R.~F.~Braga and H.~L.~Carrion,
  Phys.\ Rev.\  D {\bf 73}, 047901 (2006)
  [arXiv:hep-th/0507063].
  
\bibitem{Evans:2006ea}
  N.~Evans and A.~Tedder,
  Phys.\ Lett.\  B {\bf 642}, 546 (2006)
  [arXiv:hep-ph/0609112].
  
\bibitem{Hong:2006ta}
  D.~K.~Hong, T.~Inami and H.~U.~Yee,
  Phys.\ Lett.\  B {\bf 646}, 165 (2007)
  [arXiv:hep-ph/0609270].
  
\bibitem{Colangelo:2007pt}
  P.~Colangelo, F.~De Fazio, F.~Jugeau and S.~Nicotri,
  Phys.\ Lett.\  B {\bf 652}, 73 (2007)
  [arXiv:hep-ph/0703316].
  
\bibitem{Forkel:2007ru}
  H.~Forkel,
  Phys.\ Rev.\  D {\bf 78}, 025001 (2008)
  [arXiv:0711.1179 [hep-ph]].
\bibitem{Vega:2008af}

  A.~Vega and I.~Schmidt,
  Phys.\ Rev.\  D {\bf 78}, 017703 (2008)
  [arXiv:0806.2267 [hep-ph]].
  
\bibitem{Nawa:2008xr}
  K.~Nawa, H.~Suganuma and T.~Kojo,
  Mod.\ Phys.\ Lett.\  A {\bf 23}, 2364 (2008)
  [arXiv:0806.3040 [hep-th].
\bibitem{dePaula:2008fp}

  W.~de Paula, T.~Frederico, H.~Forkel and M.~Beyer,
  Phys.\ Rev.\  D {\bf 79}, 075019 (2009)
  [arXiv:0806.3830 [hep-ph]].
  
\bibitem{Colangelo:2008us}
  P.~Colangelo, F.~De Fazio, F.~Giannuzzi, F.~Jugeau and S.~Nicotri,
  Phys.\ Rev.\  D {\bf 78}, 055009 (2008)
  [arXiv:0807.1054 [hep-ph]].
  
\bibitem{Forkel:2008un}
  H.~Forkel and E.~Klempt,
  Phys.\ Lett.\  B {\bf 679}, 77 (2009)
  [arXiv:0810.2959 [hep-ph]].
\bibitem{Ahn:2009px}
  H.~C.~Ahn, D.~K.~Hong, C.~Park and S.~Siwach,
  Phys.\ Rev.\  D {\bf 80}, 054001 (2009)
  [arXiv:0904.3731 [hep-ph]].
  
\bibitem{Sui:2009xe}
  Y.~Q.~Sui, Y.~L.~Wu, Z.~F.~Xie and Y.~B.~Yang,
  Phys.\ Rev.\  D {\bf 81}, 014024 (2010)
  [arXiv:0909.3887 [hep-ph]].
  
\bibitem{Kapusta:2010mf}
  J.~I.~Kapusta and T.~Springer,
  Phys.\ Rev.\  D {\bf 81}, 086009 (2010)
  [arXiv:1001.4799 [hep-ph]];

\bibitem{Zhang:2010bn}
  P.~Zhang,
  Phys.\ Rev.\  D {\bf 81} (2010) 114029
  [arXiv:1002.4352 [hep-ph]];
  JHEP {\bf 1005} (2010) 039
  [arXiv:1003.0558 [hep-ph]];
  arXiv:1007.2163 [hep-ph].
  
 \bibitem{Wang:2010aj}
  See also: S.~J.~Wang, J.~Tao, X.~B.~Guo and L.~Li,
  arXiv:1007.2462 [hep-th].

\bibitem{Kwee:2007dd}
H.~J.~Kwee and R.~F.~Lebed,
  JHEP {\bf 0801}, 027 (2008)
  [arXiv:0708.4054 [hep-ph]].
  
\bibitem{Grigoryan:2007wn}
H.~R.~Grigoryan and A.~V.~Radyushkin,
  Phys.\ Rev.\  D {\bf 76}, 115007 (2007)
  [arXiv:0709.0500 [hep-ph]].
  
\bibitem{Bayona:2010bg}
  C.~A.~B.~Bayona, H.~Boschi-Filho, M.~Ihl and M.~A.~C.~Torres,
  arXiv:1006.2363 [hep-th].

  \bibitem{Brodsky:2008pg}
  S.~J.~Brodsky and G.~F.~de Teramond,
  arXiv:0802.0514 [hep-ph].

\bibitem{deTeramond:2005su}
  G.~F.~de Teramond and S.~J.~Brodsky,
  Phys.\ Rev.\ Lett.\  {\bf 94}, 201601 (2005)
  [arXiv:hep-th/0501022].

\bibitem{Grigoryan:2007my}
  H.~R.~Grigoryan and A.~V.~Radyushkin,
  Phys.\ Rev.\  D {\bf 76}, 095007 (2007)
  [arXiv:0706.1543 [hep-ph]].

\bibitem{note3} The data compilation is from
  M.~Diehl,
  Nucl.\ Phys.\ Proc.\ Suppl.\  {\bf 161}, 49 (2006)
  [arXiv:hep-ph/0510221].

 \bibitem{Grunberg}
  G.~Grunberg,
  Phys.\ Lett.\  B {\bf 95}, 70 (1980);
 Phys.\ Rev.\  D {\bf 29}, 2315 (1984);
  Phys.\ Rev.\  D {\bf 40}, 680 (1989).

   \bibitem{CSR}
  S.~J.~Brodsky and H.~J.~Lu,
  Phys.\ Rev.\  D {\bf 51}, 3652 (1995)
  [arXiv:hep-ph/9405218];
  S.~J.~Brodsky, G.~T.~Gabadadze, A.~L.~Kataev and H.~J.~Lu,
  Phys.\ Lett.\  B {\bf 372}, 133 (1996)
  [arXiv:hep-ph/9512367].

\bibitem{Appelquist:1977tw}
  T.~Appelquist, M.~Dine and I.~J.~Muzinich,
  Phys.\ Lett.\  B {\bf 69}, 231 (1977).

  \bibitem{BjorkenSR}
  J.~D.~Bjorken,
  Phys.\ Rev.\  {\bf 148}, 1467 (1966).

\bibitem{Deur:2009hu}
  A.~Deur,
  arXiv:0907.3385 [nucl-ex].

\bibitem{Brodsky:2010ur}
  S.~J.~Brodsky, G.~F.~de Teramond, A.~Deur,
  Phys.\ Rev.\  {\bf D81}, 096010 (2010)
  [arXiv:1002.3948 [hep-ph]].

\bibitem{Cornwall:1981zr}
  J.~M.~Cornwall,
  Phys.\ Rev.\  D {\bf 26}, 1453 (1982).

    \bibitem{GDH}
  S.~D.~Drell and A.~C.~Hearn,
  Phys.\ Rev.\ Lett.\  {\bf 16}, 908 (1966);
  S.~B.~Gerasimov,
  Sov.\ J.\ Nucl.\ Phys.\  {\bf 2} (1966) 430
  [Yad.\ Fiz.\  {\bf 2} (1966) 598].

\bibitem{Gao:2004zh}
  H.~Gao and L.~Zhu,
  AIP Conf.\ Proc.\  {\bf 747}, 179 (2005)
  [arXiv:nucl-ex/0411014].

\bibitem{S-Deq.}
  J.~C.~R.~Bloch,
  Phys.\ Rev.\  D {\bf 66}, 034032 (2002)
  [arXiv:hep-ph/0202073];
  P.~Maris and P.~C.~Tandy,
  Phys.\ Rev.\  C {\bf 60}, 055214 (1999)
  [arXiv:nucl-th/9905056];
  C.~S.~Fischer and R.~Alkofer,
  Phys.\ Lett.\  B {\bf 536}, 177 (2002)
  [arXiv:hep-ph/0202202];
  C.~S.~Fischer, R.~Alkofer and H.~Reinhardt,
  Phys.\ Rev.\  D {\bf 65}, 094008 (2002)
  [arXiv:hep-ph/0202195];
  R.~Alkofer, C.~S.~Fischer and L.~von Smekal,
  Acta Phys.\ Slov.\  {\bf 52}, 191 (2002)
  [arXiv:hep-ph/0205125];
  M.~S.~Bhagwat, M.~A.~Pichowsky, C.~D.~Roberts and P.~C.~Tandy,
  Phys.\ Rev.\  C {\bf 68}, 015203 (2003)
  [arXiv:nucl-th/0304003].

\bibitem{Burkert-Ioffe}
  V.~D.~Burkert and B.~L.~Ioffe,
  Phys.\ Lett.\  B {\bf 296}, 223 (1992);
  J.\ Exp.\ Theor.\ Phys.\  {\bf 78}, 619 (1994)
  [Zh.\ Eksp.\ Teor.\ Fiz.\  {\bf 105}, 1153 (1994)].

\bibitem{Parisi:1972zy}
  G.~Parisi,
  Phys.\ Lett.\  B {\bf 39}, 643 (1972).

\bibitem{Brodsky:2002cx}
  S.~J.~Brodsky, D.~S.~Hwang and I.~Schmidt,
  Phys.\ Lett.\  B {\bf 530}, 99 (2002)
  [arXiv:hep-ph/0201296].

\bibitem{Collins:2002kn}
  J.~C.~Collins,
  Phys.\ Lett.\  B {\bf 536}, 43 (2002)
  [arXiv:hep-ph/0204004].

\bibitem{Boer:2002ju}
  D.~Boer, S.~J.~Brodsky and D.~S.~Hwang,
  Phys.\ Rev.\  D {\bf 67}, 054003 (2003)
  [arXiv:hep-ph/0211110].

\bibitem{Brodsky:1995ds}
  S.~J.~Brodsky, A.~H.~Hoang, J.~H.~Kuhn and T.~Teubner,
  Phys.\ Lett.\  B {\bf 359}, 355 (1995)
  [arXiv:hep-ph/9508274].

\bibitem{Collins:2007nk}
  J.~Collins and J.~W.~Qiu,
  Phys.\ Rev.\  D {\bf 75}, 114014 (2007)
  [arXiv:0705.2141 [hep-ph]].





\end{thebibliography}
\end{document}